\documentstyle[12pt,amsfonts]{article}
\parskip=\medskipamount

\newcommand {\cA}{{\cal A}}

\newcommand {\cD}{{\cal D}}

\newcommand {\cF}{{\cal F}}
\newcommand {\cG}{{\cal G}}

\newcommand {\cL}{{\cal L}}

\newcommand {\cQ}{{\cal Q}}

\newcommand {\cV}{{\cal V}}
\newcommand {\cW}{{\cal W}}

\newcommand {\cY}{{\cal Y}}
\newcommand {\cZ}{{\cal Z}}
\newcommand{\bD}{{\bf D}}

\newcommand{\hm}{{\hat m}}
\newcommand{\hn}{{\hat n}}
\newcommand{\re}[1]{(\ref{#1})}
\def\a{\alpha}
\def\b{\beta}
\def\c{\chi}
\def\d{\delta}
\def\e{\epsilon}

\def\g{\gamma}
\def\G{\Gamma}

\def\j{\psi}
\def\k{\kappa}
\def\l{\lambda}

\def\o{\omega}

\def\q{\theta}
\def\r{\rho}
\def\s{\sigma}
\def\t{\tau}

\def\x{\xi}
\def\z{\zeta}
\def\D{\Delta}
\def\F{\Phi}
\def\J{\Psi}

\newcommand{\ad}{{\dot{\alpha}}}
\newcommand{\bd}{{\dot{\beta}}}
\newcommand{\ve}{\varepsilon}
\newcommand{\cDB}{{\bar\cD}}

\newcommand{\pa}{\partial}
\newcommand{\hf}{\frac12}

\newcommand{\Vev}[1]{\langle #1 \rangle}

\newcommand{\sect}[1]{\setcounter{equation}{0}\section{#1}}

\newcommand{\be}{\begin{equation}}
\newcommand{\ee}{\end{equation}}
\newcommand{\bea}{\begin{eqnarray}}
\newcommand{\eea}{\end{eqnarray}}
\newcommand{\non}{\nonumber}

\begin{document}
\begin{titlepage}
\thispagestyle{empty}

\begin{flushright}
LAPTH-684/98\\ JINR E2-98-54 \\ ITP-UH-08/98\\ hep-th/9805152 \\
\end{flushright}

\begin{center}
\large{{\bf N = 2 Rigid Supersymmetry with Gauged Central Charge} }
\vspace{1cm}

\large{N. Dragon${}^{\dag}$, E. Ivanov${}^{\ddag}$,
S. Kuzenko${}^{\dag ,}$\footnote{ Alexander von Humboldt
Research Fellow. On leave from Department of Physics,
Tomsk State University, Tomsk 634050, Russia
(address after February 1, 1998); E-mail: kuzenko@phys.tsu.ru.},
E. Sokatchev${}^{\P}$ and U. Theis${}^{\dag}$
}
\vspace{3mm}

${}^{\dag}$ \footnotesize{{\it Institut f\"ur Theoretische Physik,
Universit\"at Hannover\\
Appelstra{\ss}e 2, 30167 Hannover, Germany}\\
\tt{dragon,kuzenko,utheis@itp.uni-hannover.de} }
\vspace{2mm}

${}^{\ddag}$ \footnotesize{{\it Bogoliubov Laboratory of Theoretical
Physics,
Joint Institute for Nuclear Research \\
141980 Dubna, Russia}\\
\tt{eivanov@thsun1.jinr.ru}}
\vspace{2mm}

${}^{\P}$ \footnotesize{{\it Laboratoire de Physique Th{\'e}orique LAPTH
\footnote{URA 14-36 du CNRS, associ{\'e}e {\`a} l'Universit{\'e} de Savoie.}\\
LAPP, BP 110, F-74941  Annecy-le-Vieux Cedex, France }\\
\tt{sokatche@lapp.in2p3.fr} }
\end{center}

\begin{abstract}
We develop a general setting for $N=2$ rigid supersymmetric field theories with
gauged central charge in harmonic superspace. We consider those $N=2$
multiplets which have a finite number of off-shell components and exist off
shell owing to a non-trivial central charge. This class includes, in
particular, the hypermultiplet with central charge and various versions of the
vector-tensor multiplet. For such theories we present a manifestly
supersymmetric universal action. Chern-Simons couplings to an external $N=2$
super Yang-Mills multiplet are given, in harmonic superspace, for both the
linear and nonlinear vector-tensor multiplets with gauged central charge. We
show how to deduce the linear version of the vector-tensor multiplet from six
dimensions.
\end{abstract}
\vspace{5mm}


\vfill
\null
\end{titlepage}

\newpage
\setcounter{page}{1}
\setcounter{footnote}{0}

\sect{Introduction}

Supersymmetric theories with gauged central charge were introduced for the
first time in the context of $N=2$ supergravity \cite{z,dvv}. Recently, there
was a revival of interest in such theories, mainly because of the conjecture
\cite{wkll} that one of the important $N=2$ multiplets with a non-trivial
central charge, the vector-tensor (VT) multiplet \cite{ssw}, describes the
dilaton-axion complex in heterotic $N=2$ four-dimensional supersymmetric
string vacua. In particular, it was found that, besides the original `linear'
version of this multiplet, there exists its new `nonlinear' version
\cite{claus1}. Chern-Simons couplings of both versions to external $N=2$
vector multiplets \cite{claus2} and $N=2$ supergravity \cite{claus3} were
constructed. One of the important observations made in \cite{claus2} is that
in the case of the linear VT multiplet with gauged central charge for
ensuring the rigid scale and chiral invariances of the action one needs at
least one extra background abelian vector multiplet in addition to that
associated with the central charge. No such extension is required in the case
of the non-linear VT multiplet. The scale and chiral invariances are of
crucial importance for a self-consistent coupling of the VT multiplet to
conformal $N=2$ supergravity.

All these studies were carried out in the component field approach making use
of the superconformal multiplet calculus. The supersymmetry transformation
laws and the invariant actions look rather complicated in such an approach.
In this connection, the authors of \cite{claus3} noticed: ``...the complexity
of our results clearly demonstrates the need for a suitable superspace
formulation.''

Superspace formulations for the linear version of the VT multiplet with rigid
central charge were constructed in $N=2$ global central charge superspace
\cite{how,ghh,bho} and in $N=2$ harmonic superspace \cite{dkt}. The latter
approach seems to be most advantageous, since the $N=2$ harmonic superspace
\cite{gikos} provides a universal framework for general $N=2$ matter and
super Yang-Mills theories, as well as $N=2$ supergravity. The nonlinear VT
multiplet with rigid central charge was formulated in harmonic superspace in
\cite{dk,is}. In \cite{is} a new version of the nonlinear VT multiplet was
proposed in which the component vector field (the field strength of an
antisymmetric tensor field in the old version) cannot be expressed in terms
of a potential.

First steps toward superspace formulations of globally $N=2$ supersymmetric
theories with gauged central charge have been undertaken in \cite{dt}. In
particular, the $N=2$ superspace description of the linear VT multiplet with
gauged central charge was given for the first time.

In the present paper we develop a general formalism for globally $N=2$
supersymmetric theories with gauged central charge in the framework of
harmonic superspace. Our study should be considered as a preparatory step on
the way to the full local case with couplings to $N=2$ supergravity. We
propose a supersymmetric action which reproduces the component action given
in \cite{dvv}. We describe the superfield formulations of the linear and the
`old' nonlinear VT multiplet with gauged central charge. We present the
harmonic superspace form of the Chern-Simons couplings to an external $N=2$
super Yang-Mills multiplet for both the linear and nonlinear VT multiplets
with gauged central charge. Finally, we show how all the results concerning
the linear VT multiplet can be obtained by dimensional reduction from six
dimensions.

We would like to point out that one should distinguish two classes of $N=2$
multiplets with central charge. One of them is described by constrained
superfields with a finite number of off-shell components. The constraints do
not put the superfields on shell only owing to the presence of a non-trivial
central charge. This class includes the Fayet-Sohnius hypermultiplet
\cite{fs,fs1} in the off-shell version of ref. \cite{fs1} and all versions of
the VT multiplet. Our consideration here is limited just to this class.
Another type of $N=2$ multiplets is represented by unconstrained analytic
harmonic superfields with infinitely many auxiliary components like the
universal $q^+$ hypermultiplet \cite{gikos}. One can introduce and gauge the
central charge for such superfields too, but its presence or absence has no
impact on the relevant off-shell content.

The paper is organized as follows. In Section \ref{fund} we give a general
discussion of $N=2$ supersymmetric theories with gauged central charge and
review the harmonic superspace technique \cite{gikos} in the form adapted to
our study. We derive the manifestly supersymmetric action underlying the
dynamics of $N=2$ theories with gauged central charge. We also discuss the
supersymmetry transformations of the component fields and the limit of rigid
central charge. In Section \ref{lin} we start by reviewing the Fayet-Sohnius
hypermultiplet and the linear VT multiplet with gauged central charge.
Further, the superfield consistency conditions are derived which must be
fulfilled for any consistent VT multiplet superfield formulation. We then
construct the Chern-Simons coupling to an external $N=2$ vector multiplet for
the linear VT multiplet with gauged central charge. The model of linear VT
multiplet with scale and chiral invariance is considered in detail. In
Section \ref{nonlin} the nonlinear VT multiplet (its `old' version) with
gauged central charge is described in harmonic superspace. We also present
its Chern-Simons coupling to an external $N=2$ vector multiplet. In Section
\ref{6D} we demonstrate that the linear VT multiplet constraints described in
the preceding sections have a natural origin in six dimensions.

\sect{Fundamentals} 
\label{fund}

\subsection{Preliminaries} 

The theories exhibiting invariance under rigid $N=2$ supersymmetry and local
central charge transformations can be formulated in $N=2$ superspace ${\Bbb
R}^{4|8}$ with coordinates
\be
z^M=(x^m,\theta_i^\alpha,
\bar \theta^i_{\dot{\alpha}}) \qquad \overline{\theta_i^\alpha}=\bar
\theta^{{\dot{\alpha}}\,i} \qquad i=1,2.
\ee
The basic objects of these formulations are gauge covariant derivatives
\be
{\cal D}_M \equiv({\cal D}_m,{\cal D}^i_\alpha,
\bar {\cal D}^{\dot{\alpha}}_i)= D_M +
\cA_M \D \,,
\label{1}
\ee
where $D_M = (\pa_m, D^i_\a, {\bar D}^\ad_i)$
are the flat covariant derivatives
\be
\{ D^i_\a , D^j_\b\} = \{ {\bar D}_{\ad i}, {\bar D}_{\bd j} \} = 0 \qquad
\{ D^i_\a , {\bar D}_{\ad j} \} = - 2 {\rm i} \, \d^i_j \, \pa_{\a \ad}~,
\ee
$\D$ is the generator of central charge transformations (the mass dimension
of $\D$ is chosen to be $+1$) and $\cA_M (z)$ is the corresponding superfield
gauge connection. The generator $\D$ is often interpreted as the derivative
with respect to an extra bosonic coordinate $x^5$. In Section \ref{6D} we
shall explain its origin from a six-dimensional point of view. For the time
being, we simply require $\D$ to obey the Leibniz rule along with the
conditions
\be
[ \D, D_M ] = [ \D, \cD_M ] = 0\,.
\ee
The covariant derivatives are required to satisfy the constraints
\bea
\{ \cD^i_\a , \cD^j_\b\} = - 2\ve_{\a \b} \ve^{ij} {\bar \cZ} \D \quad &
\quad & \quad
\{ {\bar \cD}_{\ad i}, {\bar \cD}_{\bd j} \} =
- 2\ve_{\ad \bd} \ve_{ij}  \cZ \D \non \\
\{ \cD^i_\a ,  {\bar \cD}_{\ad j} \} & = & - 2{\rm i}
\, \d^i_j \,\cD_{\a \ad}
\label{cda}
\eea
which mean that $\cA_M$ describe an abelian vector multiplet \cite{gsw}.
The superfield strengths $\cZ$, ${\bar \cZ}$ obey the Bianchi identities
\be
{\bar \cD}_{\ad i} \cZ = 0 \qquad
\cD^{\a ( i} \cD_\a^{j)} \cZ = {\bar \cD}_\ad^{(i} {\bar \cD}^{j ) \ad}
{\bar \cZ} \,.
\label{bianchi}
\ee

Local central charge transformations are realized on $\cD_M$ and matter
superfields $U$ as
\bea
\d \cD_M & = & [\t \D, \cD_M ] \quad \Longleftrightarrow \quad
\d \cA_M = - D_M \t \non \\
\d U & = & \t \D U \equiv \t U^{(\D)}~,
\label{tau}
\eea
with $\t = \t (z)$ being an unconstrained real gauge parameter. The
interpretation of $\D$ as the derivative in an extra bosonic coordinate is
useful for recognizing the fact that $U^{(\D)}$ cannot be represented, in
general, as a linear combination of original superfields $U$ and their
covariant derivatives. Applying the central charge transformations to
$U^{(\D)}$ leads to new superfields $U^{(\D \D)}$, and so on and so forth.
Generally, in a supersymmetric theory we have some set of basic (`primary')
superfields $U$ and an infinite tower of descendants $\{ U^{(\D)}, U^{(\D
\D)},\cdots \}$. If we wish to have an off-shell multiplet with a finite
number of component fields, the superfields should be subject to some
gauge-covariant constraints. The r\^ole of such constraints is not only to
express the descendants in terms of a few basic superfields, but also to
completely specify the central charge transformations. We will consider the
case of a single central charge, though $N=2$ supersymmetry in general admits
two such charges (that would amount to a complex $\D$). The main motivation
for this restriction is just the desire to have a finite multiplet; in the
presence of two central charges the sequence $\{ U, U^{(\D)}, U^{(\D \D)},
\cdots \}$ in many cases does not terminate at any finite step (for an
alternative explanation see Section \ref{6D}, where the two central charges
are interpreted as two extra coordinates $x^{5,6}$).

It is worth saying that the superalgebra \re{cda}, as it stands, looks like
the algebra of covariant derivatives of some abelian $N=2$ gauge theory with
an unspecified gauge generator $\D$. The interpretation of the latter as a
central charge generator necessarily requires non-zero background values of
the gauge connections corresponding to $\Vev{\cZ} = {\rm const} \neq 0$. Just
with this choice of the flat limit the algebra \re{cda} goes into that of the
covariant spinor derivatives corresponding to $N=2$ supersymmetry with $\D$
as the rigid central charge. This issue will be discussed in more detail in
subsection \ref{susytrans}.

\subsection{Harmonic superspace formulation} 

The main virtue of the harmonic superspace method \cite{gikos} consists in
providing the unique possibility to describe the off-shell $N=2$
hypermultiplets in terms of unconstrained superfields. Also, it results in
off-shell formulations of $N=2$ gauge theories and supergravity in terms of
the connections and vielbeins covariantizing the derivatives with respect to
harmonic variables. The formulation of the vector gauge multiplet in the
standard $N=2$ superspace ${\Bbb R}^{4|8}$ turns out to be a gauge-fixed
version of that in $N=2$ harmonic superspace \cite{gikos}.

The harmonic superspace is also indispensable for formulating $N=2$ theories
with gauged central charge in a manifestly supersymmetric way. Since our
consideration will be essentially based upon the harmonic superspace
techniques, we start by recalling some salient features of this method.

The $N=2$ harmonic superspace is an extension of ordinary $N=2$ superspace
${\Bbb R}^{4|8}$ by the harmonic variables ${u_i}^-\,,\,{u_i}^+$ which
parametrize the two-sphere ${\rm S}^2 ={\rm SU}(2)/{\rm U}(1)$, SU(2) being
the automorphism group of $N=2$ supersymmetry,
\bea
& ({u_i}^-\,,\,{u_i}^+) \in {\rm SU}(2)\non\\
& u^+_i = \ve_{ij}u^{+j} \qquad \overline{u^{+i}} = u^-_i
\qquad u^{+i}u_i^- = 1 \;.
\eea
Tensor fields over ${\rm S}^2$ are in a one-to-one correspondence with
functions on SU(2) possessing definite harmonic U(1)-charges. A function
$\Psi^{(p)}(u)$ is said to have the harmonic U(1)-charge $p$ if
$$
\Psi^{(p)}({\rm e}^{{\rm i}\a} u^+,{\rm e}^{-{\rm i}\a} u^-)=
{\rm e}^{{\rm i}\a p} \Psi^{(p)}(u^+,u^-) \qquad |{\rm e}^{{\rm i}\a}| =
1\;.
$$
Such functions extended to the whole harmonic superspace ${\Bbb R}^{4|8}
\times {\rm S}^2$, that is $\Psi^{(p)}(z,u)$, are called harmonic $N=2$
superfields.

The operators
\bea
& D^{\pm \pm} =u^{\pm i} \pa / \pa u^{\mp i} \qquad
D^0 = u^{+i} \pa / \pa u^{+i} - u^{-i} \pa / \pa u^{-i} \non
\\
& [D^0,D^{\pm \pm}] = \pm 2D^{\pm \pm} \qquad [D^{++}, D^{--}] = D^0
\eea
are left-invariant vector fields on SU(2). $D^{\pm \pm}$ are two
independent harmonic covariant derivatives on ${\rm S}^2$, while $D^0$ is
the U(1) charge operator, $D^0 \Psi^{(p)} = p \Psi^{(p)}$.

Using the harmonics, one can convert the spinor covariant derivatives into
SU(2)-invariant operators on ${\Bbb R}^{4|8} \times {\rm S}^2$
\be
{\cal D}^\pm_\alpha = {\cD}^i_\alpha u^\pm_i
\qquad {\bar{\cal D}}^\pm_{\dot\alpha}={\bar{\cD}}^i_{\dot\alpha} u^\pm_i
\;.
\ee
Then the superalgebra (\ref{cda}) implies the existence of the anticommuting
subset
\be
\{\cD^+_\a,\cD^+_\b\} = \{\cDB^+_ {\dot\a},
\cDB^+_{\dot\b}\}= \{\cD^+_\a, \cDB^+_{\dot\a} \}=0~,
\label{zero}
\ee
whence
\bea
&& D^+_\a \cA_\b + D^+_\b \cA_\a = D^+_\a {\bar \cA}_{\dot\b} +
{\bar D}^+_ {\dot\b}\cA_\a =
0 \quad \Rightarrow \quad (\cA_\b, \;{\bar \cA}_{\dot\b}) = (D^+_\a \cG, \;
{\bar D}^+_\ad \cG) \nonumber \\
&& \cD^+_\a = D^+_\a + (D^+_\a \cG) \D~, \qquad
\cDB^+_{\dot\a}= {\bar D}^+_\ad + ({\bar D}^+_\ad \cG) \D\;.
\label{bridge}
\eea
Here the superfield $\cG = \cG (z,u)$, called `bridge', has vanishing
harmonic U(1)-charge and is real, $\breve{\cG} = \cG$, with respect to
the generalized conjugation $\; \breve{} \;\equiv \;\stackrel{\star}{\bar{}}$
\cite{gikos}, where the operation ${}^\star$ is defined by
$$
(u^+_i)^\star = u^-_i~, \quad
(u^-_i)^\star = - u^+_i, \quad \Rightarrow \quad
(u^{\pm}_i)^{\star \star} = - u^{\pm}_i.
$$

Eq.\ (\ref{bridge}) solves the constraints (\ref{cda}) under some additional
restriction on $\cG (z,u)$ (see below). An obvious immediate consequence of
the relations (\ref{zero}) and (\ref{bridge}) is the existence of an
important subclass of harmonic superfields, the covariantly analytic ones.
They are defined by the constraints
\be  \label{anconstr}
\cD^+_\a\F^{(p)}=\cDB^+_{\dot\a} \F^{(p)}=0~,
\ee
whence
\be
\F^{(p)}={\rm e}^{-\cG \D}\,\hat{\F}^{(p)} \qquad
D^+_\a \hat{\F}^{(p)}={\bar D}^+_{\dot\a} \hat{\F}^{(p)}=0\;. \label{ltrel}
\ee
The superfields $\hat{\F}^{(p)}$ are functions over the so-called analytic
subspace of the harmonic superspace parameterized by
\be  \label{analytsub}
\{\z, u^{\pm}_i \} \equiv\{
x^m_A,\q^{+\a},{\bar\q}^+_{\dot\a}, u^\pm_i \}~, \qquad \hat{\F}^{(p)}
\equiv  \hat{\F}^{(p)} (\z, u)~,
\ee
where \cite{gikos}
\be
x^m_A = x^m - 2{\rm i} \q^{(i}\s^m {\bar \q}^{j)}u^+_i u^-_j \qquad
\q^\pm_\a = \q^i_\a  u^\pm_i
\qquad {\bar \q}^\pm_{\dot\a}={\bar
\q}^i_\ad u^\pm_i\;.
\label{analbasis}
\ee
That is why such superfields are called analytic. Note that the analytic
subspace \re{analytsub} is closed under $N=2$ supersymmetry transformations
and the generalized conjugation ``$\; \breve{} \;$'', i.e.\ it is real with
respect to this conjugation. The analytic superfields with even U(1) charge
can so be chosen real.

In accordance with eq.\ \re{bridge}, the bridge possesses a more general
gauge freedom than the original harmonic-independent $\t$-group \re{tau}:
\be
\d \cG = \l - \t
\qquad \l=\l(\z,u) \qquad
D^+_\a \l = {\bar D}^+_\ad \l=0 \;.
\ee
Here the unconstrained analytic gauge parameter $\l$ has vanishing
U(1)-charge and is real, $\breve{\l} = \l$, with respect to the
analyticity preserving conjugation. The set of all $\l$-transformations is
called the $\l$-group. The $\t$-group acts on $\Phi^{(p)}$ and leaves
$\hat{\F}^{(p)}$ unchanged; the $\l$-group acts on $\hat{\F}^{(p)}$ as
\be
\d \hat{\F}^{(p)}= \l \D \hat{\F}^{(p)}
\ee
and leaves $\Phi^{(p)}$ unchanged. Thus one can equivalently formulate
the theory in the two frames related by the similarity operator
${\rm e}^{-\cG \D}$: in the $\t$-frame where the $\t$-group is manifest
and in the $\l$-frame with the $\l$-group manifest.

As we observe from the relation \re{ltrel}, the $N=2$ harmonic analyticity
is covariant in the $\t$-frame, but it becomes manifest in the $\l$-frame.
This can also be seen by comparing the covariant derivatives in both frames.

In the $\t$-frame, the complete set of gauge-covariant derivatives reads
\be
\cD_{\underline{M}}\equiv(\cD_M,\cD^{++},\cD^{--},
\cD^0) \qquad \cD^{\pm\pm}=D^{\pm\pm} \qquad
\cD^0=D^0
\ee
and their transformation law is the same as that of $\cD_M$ given by
\re{tau}. The transformation law of matter superfields $U$ is also given by
\re{tau}.

In the $\l$-frame, the covariant derivatives
\be
\hat{\cD}_{\underline{M}} =
{\rm e}^{\cG \D} \cD_{\underline{M}} {\rm e}^{-\cG \D} =
\cD_{\underline{M}} -
( D_{\underline{M}} \cG ) \D
\ee
and the matter superfields
\be
\hat{U} = {\rm e}^{\cG \D} U
\ee
transform by the rule
\be  \label{transfl}
\d \hat{\cD}_{\underline{M}} =  [\l \D ,\hat{\cD}_{\underline{M}} ]
\qquad \d \hat{U} = \l \D \,\hat{U}\;.
\ee
In this frame we have
\bea
& {} & \hat{\cD}^+_\a=D^+_\a \qquad \hat{\overline{\cD}}{}^+_{\dot\a}=
{\bar D}^+_{\dot\a} \qquad \hat{\cD}^0 = D^0 \non \\
& {} & \hat{\cD}^{\pm\pm}= D^{\pm\pm} - (D^{\pm\pm} \cG) \D
\equiv D^{\pm \pm} + \cV^{\pm \pm} \D \;.
\label{lcd}
\eea
We observe that the $u^+$-projections of the spinor covariant derivatives
contain no central charge gauge connections in the $\l$-frame, so the
conditions \re{anconstr} imply exact harmonic analyticity (after passing
to the analytic basis in ${\Bbb R}^{4|8}\times {\rm S}^2$ according to eq.\
\re{analbasis}, these projections become partial derivatives in $\q^-_\a,
\bar \q^-_{\dot\a}$, and conditions \re{anconstr} simply mean independence
of these coordinates).

The algebra of covariant derivatives, with all the isospinor indices
converted into U(1) ones, clearly does not depend on the choice of the
frame and/or the basis in ${\Bbb R}^{4|8}\times {\rm S}^2$. It reads
\bea
& {} & \qquad \qquad \{ {\bar \cD}{}^+_{\dot\a},\cD^-_\a\}=
-\{ \cD^+_\a , {\bar \cD}{}^-_{\dot\a}\}=
2{\rm i}\cD_{\a{\dot\a}} \non \\
& {} &\{ \cD^+_\a,\cD^-_\b\}=2\ve_{\a\b}
{\bar \cZ} \,\D\qquad  \quad \quad
\{ {\bar \cD}{}^+_{\dot\a},
{\bar \cD}{}^-_{\dot\b}\}=-2
\ve_{{\dot\a}{\dot\b}} \cZ \,\D \non \\
& {} & [\cD^{\pm \pm},\cD^\mp_\a]
=\cD^\pm_\a \qquad \qquad \qquad
[\cD^{\pm \pm},
{\bar \cD}{}^\mp_{\dot\a}]=
{\bar \cD}{}^\pm_{\dot\a}
\non \\
& {} &[\cD^0, \cD^{\pm\pm}] =
\pm 2 \cD^{\pm\pm} \qquad \qquad
[ \cD^{++}, \cD^{--}]= \cD^0 \;.
\label{cda2}
\eea
All other (anti-)commutators vanish except those involving the vector
covariant derivatives. The latter can be readily derived from the relations
given above.

Consideration in the $\l$-frame allows one to reveal the basic unconstrained
object of the theory. This is the connection $\cV^{++}$ covariantizing the
harmonic derivative $\cD^{++}$ with respect to the $\l$-group.

Since $[\hat{\cD}^{++},\hat{\cD}^+_\a] = [\hat{\cD}^{++},
\hat{\overline{\cD}}{}^+_\ad] = 0$, the connection $\cV^{++}$ is an analytic
real superfield, $D^+_\a \cV^{++} = {\bar D}^+_\ad \cV^{++} =0$,
$\breve{\cV}^{++}=\cV^{++}$, with the transformation law
\be
\d \cV^{++} = - D^{++}\lambda\;.
\label{astl}
\ee
No other constraints on $\cV^{++}$ emerge.

To demonstrate that all other central-charge connections are expressed
through the single object $\cV^{++}$, one should firstly solve for
$\cV^{--}$ the zero-curvature condition
\be\label{prisvoim}
D^{++} \cV^{--} - D^{--} \cV^{++} = 0\;,
\ee
which follows from considering the last commutator of harmonic derivatives
in \re{cda2} in the $\l$-frame. Its solution exists and can be given
explicitly \cite{zup}. Then, using the remainder of the
(anti-)commutation relations (\ref{cda2}) as well as the explicit
form \re{lcd} of $\hat{\cD}^+$ and $\hat{\overline{\cD}}{}^+$, one easily
expresses all the remaining connections in $\hat{\cD}_M$ in terms of
$\cV^{--}$ and, hence, $\cV^{++}$. The superfield strengths are expressed
through $\cV^{--}$ as
\be  \label{strV}
\cZ = - \frac{1}{4} (\bar D^+)^2 \cV^{--} \qquad
\bar \cZ = - \frac{1}{4} (D^+)^2 \cV^{--}\;.
\ee
The Bianchi identities \re{bianchi} can be shown to be identically satisfied.
Note that the second identity can be written as
\be
(D^+)^2 \cZ = (\bar D^+)^2\bar \cZ \label{Ban}~.
\ee

Thus, $\cV^{++}$ is indeed the fundamental unconstrained analytic
prepotential of the theory. Note that the defining relation $\cV^{++}
= - D^{++}\cG $ should be treated as a constraint on the bridge $\cG$
serving to express $\cG$ through $\cV^{++}$.

To be convinced that $\cV^{++}$ accommodates the standard component fields
content of the $N=2$ abelian vector multiplet, one should make use of the
gauge freedom \re{astl} and pass to the Wess-Zumino (WZ) gauge
\bea
\cV^{++}(\z, u)
& = & (\q^+)^2{\bar Z}(x_A)
+ ({\bar \q}^+)^2 Z(x_A)
-2 {\rm i}\q^+ \s^m {\bar \q}^+ A_m(x_A) \non \\
& {} & -2 ({\bar \q}^+)^2 \q^{+\a} \J^i_\a (x_A) u^-_i
+2 (\q^+)^2 {\bar \q}^+_\ad {\bar \J}^{i \ad}(x_A) u^-_i\non\\
& {} & + 3 (\q^+)^2 ({\bar \q}^+)^2  Y^{(ij)}(x_A) u^-_i u^-_j~.
\label{wz}
\eea
Here, $A_m$ is a real vector field, and $\J^i_\a$ and $Y^{(ij)}$ satisfy
the reality conditions
\be
\overline{\J^i_\a} = {\bar \J}_{i \ad} \qquad
\overline{Y^{(ij)}} = Y_{(ij)}\;.
\ee
The residual gauge freedom is given by $\l = \x(x_A)$ describing the
ordinary central charge transformations, with $A_m$ being the corresponding
gauge field. In the WZ gauge, $\cV^{--}$ reads
\bea
& {} &
\cV^{--}(\z, \q^-, {\bar \q}^-, u)
= (\q^-)^2{\bar Z}(x_A)
+ ({\bar \q}^-)^2 Z(x_A)
-2 {\rm i}\q^- \s^m {\bar \q}^- A_m(x_A) \non \\
& {} & + \big\{ ({\bar \q}^-)^2 \q^-  u^+_i
- ({\bar \q}^-)^2 \q^+  u^-_i
-2 ({\bar \q}^+ {\bar \q}^-) \q^+ u^-_i \big\}  \J^i(x_A)\non \\
& {} & + \big\{ (\q^-)^2 {\bar \q}^+  u^-_i
- (\q^-)^2 {\bar \q}^-   u^+_i
+ 2 (\q^+ \q^-) {\bar \q}^- u^-_i \big\}  {\bar \J}^i(x_A) \non \\
& {} & + \big\{ (\q^-)^2 ({\bar \q}^+)^2  u^-_i u^-_j
+ (\q^+)^2 ({\bar \q}^-)^2  u^-_i u^-_j
+ (\q^-)^2 ({\bar \q}^-)^2  u^+_i u^+_j \big\} Y^{ij}(x_A)\non \\
& {} & -2 \big\{ (\q^+ \q^-) ({\bar \q}^-)^2  u^+_i u^-_j
+ (\q^-)^2 ({\bar \q}^+ {\bar \q}^-)  u^+_i u^-_j \big\}
Y^{ij}(x_A) \non \\
& {} & +4 (\q^+ \q^-)({\bar \q}^+ {\bar \q}^-)Y^{ij}(x_A) u^-_i u^-_j
\;\; + \;\;\cdots
\eea
where the dots mean the terms involving derivatives of the fields. It is
worth keeping in mind that in the analytic basis \re{analbasis} $D^{++}$
and $D^{--}$ read
\bea
D^{++} & = & u^{+i} \frac{\pa}{\pa u^{-i}}
- 2{\rm i} \q^+ \s^m {\bar \q}^+ \frac{\pa}{\pa x_A^m}
+ \q^{+ \a} \frac{\pa}{\pa \q^{- \a}}
+ {\bar \q}^{+\ad} \frac{\pa}{\pa {\bar \q}^{-\ad}} \non \\
D^{--} & = & u^{-i} \frac{\pa}{\pa u^{+i}}
- 2{\rm i} \q^- \s^m {\bar \q}^- \frac{\pa}{\pa x_A^m}
+ \q^{- \a} \frac{\pa}{\pa \q^{+ \a}}
+ {\bar \q}^{-\ad} \frac{\pa}{\pa {\bar \q}^{+\ad}} \;.
\eea
Using the above relations, one easily computes the components of the
superfield strength $\cZ$ given by eqs.\ \re{strV}
\bea
& \cZ| = Z \qquad \cD^i_\a \cZ| = \J^i_\a \qquad
- \frac{1}{4} \cD^i \cD^j \cZ| = Y^{ij} \qquad
- \frac{1}{8} \cD_\a^i \cD_{\b i} \cZ| = F_{\a \b}\non \\
& F_{mn} = \pa_m A_n - \pa_n A_m \qquad
(\s^m)_{\a \ad} (\s^n)_{\b \bd} F_{mn} = 2 \ve_{\ad \bd} F_{\a \b}
+ 2\ve_{\a \b} {\bar F}_{\ad \bd}~.
\label{projections}
\eea
Here $U|$ denotes the $\q$-independent component of a superfield $U$.

Note that the formulation described so far actually coincides with the
standard harmonic superspace formulation of abelian $N=2$ gauge theory
\cite{gikos}. As was already mentioned in the previous subsection, the
crucial assumption allowing to interpret the theory as that of gauged central
charge amounts to specifying the appropriate flat limit characterized by a
non-zero background value of $\cZ$ and, hence, of $\cV^{++}$. We discuss this
point in the next subsection.

\subsection{Supersymmetry transformations} 
\label{susytrans}

Let us turn to a more detailed study of the supersymmetry transformations.

In setting up the invariant supersymmetric actions we will deal with the set
of the matter basic superfields $\hat{U}$, their covariant derivatives ($\D$
is to be included into the set of covariant derivatives) and the harmonic
gauge connection $\cV^{++}$. The construction presented in the previous
subsection is covariant under the local gauge transformations generated by
$\D$ and given by eqs. \re{transfl}. Obviously, it is also covariant under
the standard rigid $N=2$ supersymmetry transformations
\be
\d_\e \,\hat{U}  =
{\rm i} \Big( \e^{-\a} Q^+_\a
+ {\bar \e}^{-\ad} {\bar Q}^+_\ad
- \e^{+\a} Q^-_\a
- {\bar \e}^{+\ad} {\bar Q}^-_\ad  \Big) \hat{U}
\label{susy1}
\ee
and similar ones for $\cV^{++}$, where
\bea
Q^-_\a & = & - {\rm i}\frac{\pa}{\pa \q^{+ \a}} \qquad
Q^+_\a = {\rm i}\frac{\pa}{\pa \q^{- \a}}
+ 2(\s^m {\bar \q}^+)_\a \frac{\pa}{\pa x_A^m}
\non \\
{\bar Q}^-_\ad & = & - {\rm i}\frac{\pa}{\pa {\bar \q}^{+ \ad}} \qquad
{\bar Q}^+_\ad  = {\rm i}\frac{\pa}{\pa {\bar \q}^{- \ad}}
- 2 (\q^+ \s^m)_\ad \frac{\pa}{\pa x_A^m} \;.
\label{gen1}
\eea
The invariance under these two sets of transformations (local central charge
and rigid $N=2$ supersymmetry) will be the basic requirement to be fulfilled
by any off-shell superfield action we will deal with.

It is important to realize that these two basic invariances already imply the
invariance under another type of rigid $N=2$ supersymmetry, that with the
operator $\D$ as the central charge, provided one chooses an appropriate
`flat' background value for $\cV^{++}$.

The choice
\be \label{standlim}
\Vev{\cV^{++}}_{(1)} = 0
\ee
is clearly consistent with just the standard $N=2$ supersymmetry
\be  \label{backgr1}
Q^\pm_\a \Vev{\cV^{++}}_{(1)}  =  {\bar Q}^\pm_\ad \Vev{\cV^{++}}_{(1)} = 0~.
\ee
With this option, $\D$ never appears in the algebra of $N=2$ supersymmetry
and should be treated as some extra U(1) gauge generator.

On the other hand, one can choose a more general background
\cite{gio,bk,ikz}\footnote{In fact, the most general Ansatz consistent with the
Poincar\'e and ${\rm SU}(2)_A$ covariance reads (modulo $\l$ gauge group
transformations) $\Vev{\cV^{++}}_{(2)}  = b (\theta^+)^2  + \bar b (\bar
\theta^+)^2~,$ $b$ being a complex constant. This option can be reduced to
\re{backgr2} by a proper chiral phase rotation of $\theta^\pm_\alpha,
\bar\theta^\pm_{\dot
\alpha}$ and rescaling of the generator $\D$.}
\be \label{backgr2}
\Vev{\cV^{++}}_{(2)} = {\rm i}\;[\;(\theta^+)^2 - (\bar \theta^+)^2 \;]~.
\ee
Obviously, it does not vanish under the action of the generators of the
standard $N=2$ supersymmetry. However, the result of this action can be
cancelled by an appropriate compensating transformation from the $\lambda$
group. Namely,
\re{backgr2} is stable,
\be
\hat\d_\e\, \Vev{\cV^{++}}_{(2)} = 0~,
\label{annihmod}
\ee
against the following modified $N=2$ supersymmetry transformations
\bea
& \tilde\d_\e \,\cV^{++} = \d_\e \,\cV^{++} - D^{++} \l_0 \qquad
\tilde\d_\e \,\hat{U} = \d_\e \,\hat{U} + \l_0 \D \hat{U} \non \\
& \l_0 = -2{\rm i} (\e^-\q^+ - \bar\e^-\bar \q^+) \;.
\label{rigmod}
\eea
It is easy to read off from \re{rigmod} the expressions for the modified
$N=2$ supersymmetry generators in the realization on $\hat{U}$
\bea
\tilde Q^-_\a & = & Q^-_\a~, \qquad
\tilde Q^+_\a = {\rm i}\frac{\pa}{\pa \q^{- \a}}
+ 2(\s^m {\bar \q}^+)_\a \partial_m
-2\q^+_\a \D~,
\non \\
{\bar \cQ}^-_\ad & = & {\bar Q}^-_\ad~, \qquad
{\bar \cQ}^+_\ad  = {\rm i}\frac{\pa}{\pa {\bar \q}^{- \ad}}
- 2 (\q^+ \s^m)_\ad \partial_m
- 2{\bar \q}^+_\ad \D~.
\label{genmod}
\eea
They form the $N=2$ superalgebra with $\D$ as the central charge. What
concerns the realization of this modified $N=2$ supersymmetry on $\cV^{++}$,
it can be reduced to the standard one \re{susy1} by shifting the harmonic
connection
\be
\cV^{++} = \Vev{\cV^{++}}_{(2)}+ \hat \cV^{++}~, \qquad
\tilde\d_\e \,\hat\cV^{++} = \d_\e \,\hat\cV^{++}~.
\ee
This is of course due to the fact that $\cV^{++}$ and $\hat\cV^{++}$
possess zero central charge, $\D \cV^{++} = \D \hat\cV^{++}= 0$.

Eqs.\ \re{rigmod} and \re{genmod} yield the off-shell realization of the
central-charge extended rigid $N=2$ supersymmetry in the general case of
{\it unfixed} $\l$-group gauge freedom.

In practice, for the component considerations, it is convenient to know how
rigid $N=2$ supersymmetry is realized in the WZ gauge \re{wz}. Actually, it
makes no difference from which rigid supersymmetry transformations one
starts, \re{susy1}, \re{gen1} or \re{rigmod}, \re{genmod}, since they
differ by a $\l$-gauge transformation. Let us choose, e.g., \re{susy1} and
\re{gen1}. Then, to preserve the WZ gauge, every supersymmetry transformation
should be accompanied by a special $\e$-dependent gauge transformation with
the parameter
\be
\l (\e) = - 2 (\e^- \q^+) {\bar Z} - 2 ({\bar \e}^- {\bar \q}^+) Z
+ 2 {\rm i}( \e^- \s^m {\bar \q}^+ + \q^+ \s^m {\bar \e}^- ) A_m
+ {\rm O}(\q^2)\;.
\ee
This leads to the standard transformation law for the component fields
of the vector multiplet as well as modifies eq.\ \re{susy1} by
\be
\d_{{\rm SUSY}} \,\hat{U}  =
{\rm i} \Big( \e^{-\a} \cQ^+_\a
+ {\bar \e}^{-\ad} {\bar \cQ}^+_\ad
- \e^{+\a} \cQ^-_\a
- {\bar \e}^{+\ad} {\bar \cQ}^-_\ad  \Big) \hat{U}~,
\label{susy2}
\ee
where
\bea
\cQ^-_\a & = & Q^-_\a \qquad
\cQ^+_\a = {\rm i}\frac{\pa}{\pa \q^{- \a}}
+2{\rm i} \q^+_\a {\bar Z}(x_A) \D
+ 2(\s^m {\bar \q}^+)_\a \nabla_m + {\rm O}(\q^2)
\non \\
{\bar \cQ}^-_\ad & = & {\bar Q}^-_\ad \qquad
{\bar \cQ}^+_\ad  = {\rm i}\frac{\pa}{\pa {\bar \q}^{- \ad}}
- 2{\rm i} {\bar \q}^+_\ad Z (x_A) \D
- 2 (\q^+ \s^m)_\ad \nabla_m + {\rm O}(\q^2) \non\\
\nabla_m & \equiv &
\frac{\pa}{\pa x_A^m} + A_m (x_A)\D \;.
\label{gen2}
\eea
To compute the supersymmetry transformations of the component fields, it is
sufficient to note the identity
\bea
&& \d_{{\rm SUSY}} \hat{\cD}_{M_1} \cdots
\hat{\cD}_{M_p} (\D)^q \hat{U}| \non \\
&& =
- \Big( \e^{-\a} \hat{\cD}^+_\a
+ {\bar \e}^{-\ad} \hat{\overline{\cD}}{}^+_\ad
- \e^{+\a} \hat{\cD}^-_\a
- {\bar \e}^{+\ad} \hat{\overline{\cD}}{}^-_\ad  \Big)
\hat{\cD}_{M_1} \cdots \hat{\cD}_{M_p} (\D)^q \hat{U}|\;.
\label{susy3}
\eea
Along with the relation
\be
\hat{\cD}_{M_1} \hat{\cD}_{M_2} \cdots \hat{\cD}_{M_k} \hat{U}|
= \cD_{M_1} \cD_{M_2} \cdots \cD_{M_k} U|\;,
\label{twoframes}
\ee
which holds in the above WZ gauge (and with the appropriately fixed
$\t$-gauge group freedom), this gives us a convenient practical recipe to
compute the supersymmetry transformations of component fields.

Now, let us recall eqs.\ \re{cda} and \re{projections}. It follows from
\re{susy2}, \re{gen2} that the commutator of two supersymmetry
transformations involves not only a space-time translation, but also central
charge terms proportional to $Z$ and $\bar Z$. Therefore, \re{susy2},
\re{gen2} describe both situations discussed earlier: when $Z$ is assumed
to have a zero vacuum expectation value, we get the WZ gauge-fixed form of
the standard $N=2$ supersymmetry with $\D$ being an extra gauge group
generator; on the contrary, if $Z$ possesses a non-zero vacuum expectation
value, we are left with a gauge-fixed form of rigid supersymmetry with
the central charge $\D$. The precise correspondence with eqs.\ \re{backgr2}
and \re{genmod} emerges upon the choice
\be  \label{Zcond}
\Vev{Z} = - {\rm i}\;.
\ee

As the last remark, we mention that with our choice of the flat background
the reduction to the case of rigid central charge (which corresponds to a
constant strength $\cZ$) goes by putting
\be  \label{rigidlim}
\cV^{++} = \Vev{\cV^{++}}_{(2)} =
{\rm i}\;[\;(\theta^+)^2 - (\bar \theta^+)^2 \;]~,
\ee
or, in the WZ gauge,
\be \label{rigidlim1}
Z =
\Vev{Z} = - {\rm i} \qquad (\cZ = - \bar\cZ = -{\rm i})~,
\ee
all other components of the vector multiplet being equal to zero. This
yields, in particular, the following rigid central charge form of the
covariant derivatives $\cD_\a^i$ and ${\bar \cD}_{i \ad}$
\be
{\bf D}^i_\a = \frac{\pa}{\pa \q^\a_i}
+ {\rm i} (\s^m {\bar \q}^i)_\a \pa_m
- {\rm i} \q^i_\a \,\D
\qquad
{\bar {\bf D}}_{\ad i} = - \frac{\pa}{ \pa {\bar \q}^{\ad i} }
- {\rm i} (\q_i \s^m)_\ad \pa_m
- {\rm i}{\bar \q}_{\ad i} \,\D \;.
\label{rcccd}
\ee

As follows from the above consideration, $N=2$ supersymmetric theories with
gauged central charge can be obtained from those with rigid central charge by
covariantizing the constraints which define the multiplets in the rigid case.
However, as a rule such a covariantization requires adding some non-minimal
terms in order to give rise to consistent constraints (see Sect. \ref{lin}).

\subsection{Supersymmetric action} 

Now, we are prepared to present the main result of this Section -- the action
functional rule underlying the dynamics of $N=2$ supersymmetric theories with
gauged central charge. To avoid a possible confusion, let us point out that
here and in the rest of the paper we limit our consideration to matter
superfields $U$ with a {\it finite} number of auxiliary components and
non-trivial central charge. Such superfields are necessarily constrained, so
are the relevant Lagrangian densities. Our consideration does not directly
apply to the theories with unconstrained superfields like the $q^+$
hypermultiplet \cite{gikos} having infinitely many auxiliary fields, though
non-trivial rigid and gauged central charges can be introduced in this case
too, e.g.\ via the Scherk-Schwarz mechanism \cite{SSw}.

Let $\cL^{(ij)} (z)$ be an isovector superfield which is built out of the
basic dynamical superfields ($U, \cV^{++}, \ldots $) and possesses the
following basic properties
\be
\cD^{(i}_\a \cL^{jk)} = {\bar \cD}^{(i}_\ad \cL^{jk)} = 0\;.
\label{lm1}
\ee
The same constraints written for the harmonic superfield
\be
\cL^{++} = \cL^{ij} u^+_i u^+_j
\label{lm2}
\ee
amount to the following set
\bea
& \cD^+_\a \cL^{++} = {\bar \cD}^+_\ad \cL^{++} = 0
\label{analyticity1}\\
& D^{++} \cL^{++} = 0\;.
\label{holomorphy1}
\eea
Thus, $\cL^{++}$ is covariantly analytic and bilinear in the harmonics. In
the $\l$-frame, eqs.\ \re{analyticity1} and \re{holomorphy1} turn into
\bea
& \hat{\cL}^{++} = \hat{\cL}^{++} (\z,u)
\label{analyticity2}\\
& \hat{\cD}^{++} \hat{\cL}^{++} = \Big( D^{++}
+ \cV^{++} \D \Big) \hat{\cL}^{++} = 0\;.
\label{holomorphy2}
\eea
Let us consider the integral over the analytic subspace
\be
S = \int du\,d\z^{(-4)} \cV^{++} \hat{\cL}^{++}
\label{action}
\ee
where $d\z^{(-4)}=d^4x_A d^2\q^+d^2{\bar \q}^+$ and the integration over
SU(2) is defined by \cite{gikos}
$$
\int du \; 1 = 1 \qquad \int du \, u^+_{(i_1} \ldots u^+_{i_n} u^-_{j_1}
\ldots u^-_{j_m)} = 0 \qquad n+m > 0\;.
$$
Being manifestly $N=2$ supersymmetric, functional \re{action} is also
invariant under arbitrary local central charge transformations, since its
variation
\be
\d S = \int du\,d\z^{(-4)} \left\{ -(D^{++}\l) \hat{\cL}^{++}
+ \l \cV^{++} \D \hat{\cL}^{++}\right\}
\ee
can be transformed, by integrating the first term by parts, to
\be
\d S = \int du\,d\z^{(-4)} \l \hat{\cD}^{++} \hat{\cL}^{++}
\ee
that vanishes, as a consequence of \re{holomorphy2}. In the WZ gauge \re{wz},
it is easy to reduce $S$ to components using the integration rule
$$
\int du\,d\z^{(-4)} = \frac{1}{16} \int  d^4 x \, du\,
(D^-)^2 ({\bar D}^-)^2
$$
and the relation \re{twoframes}. This gives
\bea
 S = -\frac{1}{12}  \int d^4 x &
\Big\{ & Z \cD^{\a i} \cD_\a^j \, \cL_{ij}|
+ {\bar Z} {\bar \cD}_\ad^i {\bar \cD}^{\ad j} \cL_{ij}|
 - {\rm i} A^{\a \ad} [\cD^i_\a, {\bar \cD}_\ad^j ] \cL_{ij}|\non\\
& {} & +4 \J^{\a i} \cD^j_\a \cL_{ij}|
+ 4 {\bar \J}^i_\ad {\bar \cD}^{\ad j} \cL_{ij}|
- 12 Y^{ij} \cL_{ij}| \;\;\;\; \Big\} \;.
\label{compaction}
\eea
This is exactly the general form of the component action for $N=2$
supersymmetric theories with gauged central charge, which was suggested in
\cite{dvv} (of course, the invariant action for the gauge superfield
$\cV^{++}$ itself should be added). After performing the reduction
\re{rigidlim} to the case of rigid central charge, \re{action} goes into the
supersymmetric action functional proposed in \cite{dkt}.

At this step we observe one more reason why $\cV^{++}$ should necessarily
contain a non-vanishing background part \re{backgr2}: just this part produces
correct kinetic terms for the matter superfields $U$ present in
$\hat\cL^{++}$.

The action \re{action} provides us with a universal rule for constructing
invariants in $N=2$ theories with gauged central charge and constrained
matter superfields $U$. It is reminiscent of the action rule in general
relativity, with $\cV^{++}$ being the analog of $\sqrt{-g}$ and
$\hat\cL^{++}$ of the Lagrangian density.

Note that $\D \,\cL^{++}, \D^2 \,\cL^{++}, \ldots $ satisfy the basic
requirements \re{analyticity1} and \re{holomorphy1} and so, at first
sight, their $\l$-frame images could be equally acceptable for the r\^ole of
Lagrangian densities. However, the harmonic constraint \re{holomorphy2}
implies
$$
\cV^{++}\D^n \hat\cL^{++} = - D^{++}\D^{n-1}\hat\cL^{++}~.
$$
As a consequence, all such densities, except for $\hat\cL^{++}$ itself,
produce full harmonic derivatives upon substitution into \re{action} and so
do not contribute, both in the cases of gauged and rigid central charges.
Recalling the interpretation of $\D$ as ${\partial/\partial x^5}$ (see
Section \ref{6D}), this amounts to saying that the analytic superspace
action \re{action} does not depend on $x^5$, though the Lagrangian densities
could bear such a dependence.

In the next Sections we will illustrate the general formalism given here
by several examples of $N=2$ supersymmetric theories with gauged central
charge.

\sect{Models with linear central charge transformations} 
\label{lin}

Here we consider the simplest case of the linearly realized central charge.

\subsection{Hypermultiplet with central charge} 

Let us start with gauging the central charge of the Fayet-Sohnius
hypermultiplet \cite{fs,fs1} coupled to an external $N=2$ Yang-Mills
superfield. It is described by a superfield $q_i (z)$ and its conjugate
${\bar q}^i (z)$ subject to the constraints
\be
{\Bbb D}^{(i}_\a q^{j)} = {\bar {\Bbb D}}^{(i}_\ad q^{j)} = 0
\label{fs1}
\ee
where
\be
{\Bbb D}_M = {\bf D}_M +  {\Bbb A}_M \qquad
{\Bbb A}_M = {\Bbb A}_M^a (z) T^a \qquad
[\D \, , \, {\Bbb D}_M ] = [\D \, , \, T^a ] = 0
\ee
with ${\bf D}_M$ being the rigid central charge covariant derivatives
\re{rcccd}, ${\Bbb A}_M$ the $N=2$ super Yang-Mills connection \cite{gsw},
and $T^a$ the generators of the gauge group. Eq.\ \re{fs1} implies that
only the superfields $q_i$, $q_i^{(\D)}$ and their conjugates contain
independent component fields, and the higher descendants are expressed in
terms of $q_i$, $q_i^{(\D)}$ and their conjugates. Actually, it is easy to
show that the components of $q_i^{(\D)}$ are expressed through those of
$q_i$ and $x$-derivatives of the latter, leaving us with $(8 + 8)$ component
fields off-shell.

To gauge the central charge, it is sufficient to naively covariantize the
above constraints with respect to the central charge
\be
 \cD^{(i}_\a q^{j)} = {\bar \cD}^{(i}_\ad q^{j)} = 0 \;.
\label{fs2}
\ee
Here the covariant derivatives
\be
\cD_M = D_M + \cA_M \,\D +  {\Bbb A}_M
\ee
satisfy now the algebra
\bea
\{ \cD^i_\a , \cD^j_\b\} =
- 2\ve_{\a \b} \ve^{ij} \big({\bar \cZ} \D +  {\bar \cW}\big)
\quad & \quad & \quad
\{ {\bar \cD}_{\ad i}, {\bar \cD}_{\bd j} \} =
- 2\ve_{\ad \bd} \ve_{ij} \big( \cZ \D +  \cW \big) \non \\
\{ \cD^i_\a ,  {\bar \cD}_{\ad j} \} & = & - 2{\rm i}
\, \d^i_j \,\cD_{\a \ad}
\label{ymcda}
\eea
with the Yang-Mills superfield strength $\cW = \cW^a (z) T^a$ obeying the
Bianchi identities of the form \re{bianchi}. The central charge
transformations of the component fields can be determined in the same
fashion as it has been done in \cite{dt} for the case ${\Bbb A}_M = 0$. For
the harmonic superfield $q^+ \equiv q^i u_i^+$ the constraints \re{fs2}
amount to
\be
\cD^+ q^+ = {\bar \cD}^+ q^+ = 0 \qquad
D^{++} q^+ = 0 \;.
\label{fs3}
\ee
Therefore, $q^+$ is a covariantly analytic superfield. Passing to the
$\l$-frame with respect to both the central charge and Yang-Mills gauge
groups yields $\hat{q}{}^+$ which is analytic, $\hat{q}{}^+ = \hat{q}{}^+
(\z, u)$, and obeys the constraint
\be
\big( D^{++} + \cV^{++}\,\D +  {\Bbb V}^{++} \big)\,
\hat{q}{}^+ = 0 \;.
\label{fs4}
\ee
Here ${\Bbb V}^{++}$ is the analytic prepotential associated with
${\Bbb A}_M$ \cite{gikos}. As the Lagrangian density $\cL^{++}$ we can
choose the same expression as in the case of rigid central charge \cite{dkt}
\be  \label{qL}
\cL^{++} = \hf (\, \breve{q}{}^+ \,\D \, q^+ -
\D \,\breve{q}{}^+\;q^+\,)
-{\rm i} \,m \,\breve{q}{}^+ q^+~.
\ee
It is straightforward to check that both structures in the right-hand side
of \re{qL} solve the basic constraints \re{analyticity1} and \re{holomorphy1}.
Using \re{fs4}, one can rewrite the action in the form
\be
S = - \int du \, d\z^{(-4)} \Big\{
\breve{\hat q}{}^+ (D^{++} +  {\Bbb V}^{++} ) \hat q^+
+{\rm i} \, m\,\cV^{++} \breve{\hat q}{}^+ \hat q^+ \Big\}~.
\ee

This action looks very similar to the action of the $q$-hypermultiplet with
infinitely many auxiliary components \cite{gikos}. The crucial difference
lies, however, in that the $q$-hypermultiplet is an {\it unconstrained}
analytic superfield, while the above Fayet-Sohnius $\hat q$-hypermultiplet is
still subject to the off-shell harmonic constraint \re{fs4} which reduces the
infinite tower of components appearing in the harmonic expansion of $\hat
q^+$ to the irreducible $(8 + 8)$ content. The presence of non-trivial
central charge in the harmonic derivative in \re{fs4} is crucial for keeping
the theory off-shell: putting $\D$ equal to zero immediately makes
\re{fs4} an equation of motion. Note that for the $q$-hypermultiplet
of ref.\ \cite{gikos} one can also introduce the central charge. This can be
done by the Scherk-Schwarz mechanism, identifying $\D$ with the generator of
some U(1) symmetry of the action (e.g., of U(1) subgroup of the YM group).
One can gauge such a central charge by introducing the appropriate $V^{++}$.
However, such a central charge does not lead to any reduction of the infinite
number of off-shell degrees of freedom in $q^+$; its only effect is to provide
a mass for the $q^+$ hypermultiplet (and a physical bosons potential in the
case of self-interacting $q^+$).

\subsection{Linear vector-tensor multiplet} 
\label{linvt}

Now, let us turn to gauging the central charge of the linear free VT
multiplet \cite{ssw,wkll}. It is described by a real superfield $L(z)$
subject to the constraints \cite{dkt}
\be
{\bf D}^+_\a {\bar {\bf D}}^+_\ad L
={\bf D}^{+\a} {\bf D}^+_\a L = 0\;.
\label{lvt1}
\ee
Eq.\ \re{lvt1} implies that only the superfields $L$ and $L^{(\D)}$ contain
independent component fields, while the rest of the descendants is expressed
in terms of $L$, $L^{(\D)}$. Simultaneously, \re{lvt1} defines the action of
$\D$ on $L(z)$ and thereby expresses the components of $L^{(\D)}(z)$ in terms
of those of $L(z)$. In order to be able to treat the VT multiplet on an equal
footing both in the $\t$- and $\l$-frames, it is convenient to regard $L$ in
\re{lvt1} as a general harmonic superfield, $L = L(z,u)$, and to eliminate
the dependence on the harmonics by the additional harmonic constraint
\be  \label{harmVTc}
D^{++}L = 0~.
\ee

The VT multiplet requires more delicate treatment than the hypermultiplet,
since in this case the naive minimal covariantization with respect to the
gauged central charge yields inconsistent constraints. Namely, it turns out
that such a covariantization is incompatible with an important consistency
condition following from the harmonic constraint \re{harmVTc}.

In order to demonstrate this, let us first note that \re{harmVTc}, in the
cases of both local and rigid central charges, implies \cite{is}
\be  \label{harmVTc2}
\cD^{--}L = 0~.
\ee
The simplest way to see this is to realize that $\cD^{++}$ and $\cD^{--}$ are
the raising and lowering operators of the right SU(2) group acting on the
U(1) charges of the harmonic superfields (see \re{cda2}); $L$ is
chargeless, so \re{harmVTc} as well as \re{harmVTc2} mean that it is a
singlet of this SU(2) group.

Equation \re{harmVTc2} holds irrespectively of the precise form of the
constraints \re{lvt1}, or their covariantization. It is important that the
covariant constraints involve the gauged-central-charge-covariantized spinor
derivatives $\cD^{\pm}_\alpha, \bar\cD^{+}_{\dot\alpha}$ satisfying the
algebra \re{cda2} and commuting with $\D$. Successively acting on
\re{harmVTc2} by the derivatives $\cD^+_\alpha, \bar\cD^+_{\dot\alpha}$ and
making use of this algebra, one gets a number of useful relations. Applying
$(\cD^+\cD^+)$ and $(\cD^+\cD^+ \bar\cD^+\bar\cD^+)$ yields, respectively,
\bea
0 &=& \cD^+\cD^+  \cD^{--} L =
\left[ \cD^{--} \cD^+\cD^+ - 2 \cD^-\cD^+ - 4 \bar\cZ \D \right] L
\label{centrrel} \\
0 &=& \cD^+\cD^+  \bar\cD^+\bar\cD^+\cD^{--} L  \non \\
  &=& \cD^{--} \cD^+ \cD^+ \cDB^+ \cDB^+ L +
8{\rm i}\, \cD^{\ad\a} \cD_\a^+ \cDB_{\ad}^+ L
-
2 \cD^- \cD^+ \cDB^+ \cDB^+ L - 2 \cDB^- \cDB^+ \cD^+ \cD^+ L  \non \\
&-&  4\D \big( L \cD^+ \cD^+ \cZ + 2 \cD^+ \cZ \cD^+ L
+ 2 \cDB^+ {\bar \cZ} \cDB^+ L
+ \cZ \cD^+ \cD^+ L + {\bar \cZ} \cDB^+ \cDB^+ L \big)~.
\label{consistency1}
\eea
Eq.\ \re{centrrel} defines the action of the central charge $\D$ on $L$, and,
because of the reality of $\D$, implies a kind of reality condition for
$\cD^+_\alpha L$. Eq.\ \re{consistency1} is the consistency condition
mentioned earlier. For the case of rigid central charge, $\cZ = - {\rm i}$,
this consistency condition and eq.\ \re{centrrel} are reduced to those found
in \cite{is}.

Eq.\ \re{consistency1} severely restricts the form of possible constraints on
$L$. If, for instance, we naively covariantize the constraints of the free VT
multiplet
$$
\cD_\a^+ \cDB_{\ad}^+ L = 0 \qquad \cD^+ \cD^+ L = 0\,
$$
eq.\ \re{consistency1} would give
$$
  0 = \D \big( L \cD^+ \cD^+ \cZ + 2 \cD^+ \cZ \cD^+ L
+ 2 \cDB^+ {\bar \cZ} \cDB^+ L
        \big)
$$
which is fulfilled only if either $L$ is $\D$-invariant, thus putting the
multiplet on-shell, or if $\cZ$ is a constant, which takes us back to rigid
central charge.

To find the correct set of constraints, one should start from the general
Ansatz
\bea
\cD_\a^+ \cDB_{\ad}^+ L & = & a_1 \cD_\a^+ \cZ \cDB_{\ad}^+ L - \bar{a}_1
\cDB_{\ad}^+ {\bar \cZ} \cD_\a^+ L + a_2 \cD_\a^+ \cZ \cDB_{\ad}^+ {\bar Z}
+ a_3 \cD_\a^+ L \cDB_{\ad}^+ L\, \non\\
\cD^+ \cD^+ L & = & a_4 \cD^+ \cZ \cD^+ L + a_5 \cDB^+ {\bar Z} \cDB^+ L +
a_6 \cD^+ \cD^+ \cZ + a_7 \cD^+ \cZ \cD^+ \cZ \non \\
&&  + a_8 \cDB^+ {\bar \cZ} \cDB^+ {\bar \cZ} + a_9 \cD^+ L \cD^+ L
+ a_{10} \cDB^+ L \cDB^+ L\;
\label{ansatz}
\eea
where all the coefficients are functions of $L$, $\cZ$ and $\bar \cZ$,
and $a_2$, $a_3$ must be real. These constraints have to satisfy the
obvious conditions
\begin{equation} \label{consistency2}
\cD_\a^+ \cD^+ \cD^+ L = 0
\qquad \cDB_{\ad}^+ \cD^+ \cD^+ L = \cD^+ \cD^+ \cDB_{\ad}^+ L
\end{equation}
which produce a set of homogeneous differential equations for the coefficient
functions. However, as was shown above, not all solutions turn out to be
consistent. We have to check the condition \re{consistency1} for each
solution to single out the proper constraints.

Requiring the deformed constraints to reduce to the free ones \re{lvt1} for
$\cZ = {\rm const}$, a particular solution reads \cite{dt}
\bea
\cD_\a^+ \cDB_{\ad}^+ L & = & 0 \non \\
\cD^+ \cD^+ L & = & \frac{2}{\raisebox{-1pt}{${\bar \cZ} - \cZ$}}
\Big( \cD^+ \cZ \cD^+ L
+ \cDB^+ {\bar \cZ} \cDB^+ L + \hf L \cD^+ \cD^+ \cZ \Big)\; .
\label{lvtcc}
\eea
These constraints are linear in $L$ and its derivatives, and therefore the
theory does not include any self-coupling. A superfield Lagrangian density
which meets the requirements described in Section \ref{fund}, is given by:
\begin{equation}
\cL^{++} = -\frac{{\rm i}}{4} \Big( \cD^+ L \cD^+ L
- \cDB^+ L \cDB^+ L + L \cD^+ \cD^+ L \Big)\; .
\label{lvtaction}
\end{equation}
This model has been investigated in detail in \cite{dt}.

A more direct derivation of the constraints \re{lvtcc} (and their
generalization) will be presented in Section \ref{6D}. There the consistency
condition \re{consistency1} will be automatically solved in a six-dimensional
framework.

\subsection{Linear vector-tensor
multiplet with Chern-Simons couplings} 

The linear VT multiplet coupled to an external $N=2$ super Yang-Mills
multiplet is described by a real superfield $L(z)$ constrained by
\cite{ghh,bho,dkt}
\bea
\bD^+_\a {\bar \bD}^+_\ad L & = & 0 \non \\
\bD^+ \bD^+ L & = & \frac{g}{2} \,{\rm tr}\,
\Big( ({\bar \cD}^+)^2 {\bar \cW}^2 - (\cD^+)^2 \cW^2 \Big)
\label{lvtcs1}
\eea
with $g$ a real coupling constant. The $N=2$ super Yang-Mills field strength
$\cW$ is defined as in eq.\ \re{ymcda}. The superfield redefinition
\be
L \equiv {\Bbb L} -\hf \,g \,{\rm tr}\, (\cW - \bar \cW)^2
\ee
brings the above constraints to the form
\bea
\bD^+_\a {\bar \bD}^+_\ad {\Bbb L} & = &
-g \,{\rm tr}\, \Big(\cD^+_\a \cW {\bar \cD}^+_\ad {\bar \cW} \Big) \non \\
\bD^+ \bD^+ {\Bbb L} & = & g \,{\rm tr}\,
\Big( {\bar \cD}^+ {\bar \cW} {\bar \cD}^+ {\bar \cW}\Big)
\label{standard}
\eea
given in \cite{dkt}.

A consistent deformation of the constraints \re{lvtcs1}, which corresponds
to the gauged central charge, reads
\bea
\cD_\a^+ \cDB_{\ad}^+ L & = & 0 \non \\
\cD^+ \cD^+ L & =& \frac{2}{\raisebox{-1pt}{${\bar \cZ} - \cZ$}}
\Big( \cD^+ \cZ \cD^+ L
+ \cDB^+ {\bar \cZ} \cDB^+ L + \hf L \cD^+ \cD^+ \cZ \Big)\non \\
&& + \frac{{\rm i} \,g}{\raisebox{-1pt}{${\bar \cZ} - \cZ$}} \,{\rm tr}\,
\Big( ({\bar \cD}^+)^2 {\bar \cW}^2 - (\cD^+)^2 \cW^2 \Big)
\; .
\label{lvtcscc}
\eea
As the corresponding Lagrangian density we can again choose $\cL^{++}$ given
by eq.\ \re{lvtaction}.

To solve the differential constraints on the field strengths of the vector
and antisymmetric tensor contained in the VT multiplet, we first specify the
component fields of the external super Yang-Mills strength $\cW$
\bea
& \cW| = W \qquad \cD^i_\a \cW| = \c^i_\a \qquad
- \frac{1}{4} \cD^i \cD^j \cW| = D^{ij} \non \\
& - \frac{1}{8} \cD_\a^i \cD_{\b i} \cW| = \cF_{\a \b} \qquad
\cF_{mn} = \pa_m \cA_n - \pa_n \cA_m + [\cA_m, \cA_n]~,
\label{extcomp}
\eea
as well as those of $L$
\bea
& L= L| \qquad \l_\a^i = \cD_\a^i L| \qquad
\bar{\l}_{\ad i} = \cDB_{\ad i}L| \qquad  U = \D L| \non \\
& G_{\a\b} = \frac{1}{4} [\cD_{\a i}\,,\,\cD_\b^i] L|
\qquad \bar{G}_{\ad\bd} = \overline{G_{\a\b}} \non \\
& V_{\a\ad} = -\frac{1}{4} [\cD_\a^i\, , \, \cDB_{\ad i}] L|\; .
\label{comp}
\eea

The central charge transformations now read (hereafter, we keep only the
purely bosonic contributions)
\bea
\D (I V_m - L \pa_m R) & = & \nabla^n G_{mn} \non \\
\D (I \tilde{G}_{mn} + R G_{mn} + L F_{mn}) & = & - \ve_{mnkl}
\nabla^k V^l
\eea
while the differential constraints on $V_m$ and $G_{mn}$ look as follows:
\bea
\nabla^m (I V_m - L \pa_m R) & = & \frac{1}{2} F^{mn} G_{mn} - g \pa^m\,
{\rm tr}\, \Big[ {\rm i} (W - \bar{W}) \cD_m (W + \bar{W}) \non \\
&& + 2 \ve_{mnkl} \big(\cA^n \pa^k \cA^l + \frac{2}{3} \cA^n \cA^k \cA^l
\big) \Big] \non \\
\nabla^m (I \tilde{G}_{mn} + R G_{mn} + L F_{mn}) & = & - \tilde{F}_{mn} V^m
+ 2 g \pa^m\, {\rm tr}\, \big[ {\rm i} (W - \bar{W}) \cF_{mn} \non \\
&& + (W + \bar{W}) \tilde{\cF}_{mn} \big]\; .
\eea
Here we denoted $I = {\rm Im}\; Z$ and $R = {\rm Re}\; Z$. The general
solution of the constraints in terms of a 1-form $T_m$ and a 2-form
$K_{mn}$ reads
\bea
I V_m & = & \frac{1}{2} \ve_{mnkl} (\pa^n K^{kl} - A^n \tilde{G}^{kl}) +
L \pa_m R \non \\
&& - g\, {\rm tr}\, \Big[ {\rm i} (W - \bar{W}) \cD_m (W + \bar{W}) + 2
\ve_{mnkl} \big(\cA^n \pa^k \cA^l + \frac{2}{3} \cA^n \cA^k \cA^l \big)
\Big] \non \\
I \tilde{G}_{mn} + R G_{mn} & = & \ve_{mnkl} (\pa^k T^l - A^k V^l) - L
F_{mn} \non \\
&& + 2 g\, {\rm tr}\, \big[ {\rm i} (W - \bar{W}) \cF_{mn} + (W + \bar{W})
\tilde{\cF}_{mn} \big]\; .
\eea

In principle, all the consistency conditions are satisfied for more general
constraints than those in \re{lvtcscc}:
\bea
\cD_\a^+ \cDB_{\ad}^+ L & = & 0 \non \\
\cD^+ \cD^+ L & =& \frac{2}{\raisebox{-1pt}{${\bar \cZ} - \cZ$}}
\Big( \cD^+ \cZ \cD^+ L
+ \cDB^+ {\bar \cZ} \cDB^+ L + \hf L \cD^+ \cD^+ \cZ \Big)\non \\
&& + \frac{1}{\raisebox{-1pt}{${\bar \cZ} - \cZ$}}
\Big( (\cD^+)^2 F(\cW) +
({\bar \cD}^+)^2 {\bar F}({\bar \cW}) \Big)\label{nolabel}
\eea
where $F(\cW)$ is some holomorphic function of the chiral strength $\cW$.
But then we are unable, in general, to solve the constrained vector and
antisymmetric tensor component field of $L$ in terms of gauge two- and
one-forms, respectively\footnote{Likely, it is still possible to have a
consistent dual formulation of such a more general theory in terms of a
Lagrange multiplier vector multiplet in the spirit of ref.\ \cite{is}.}.

Let us choose $\cW$ in eq.\ \re{lvtcscc} just to be $\cZ$. Then we obtain
the following consistent constraints
\bea
\cD_\a^+ \cDB_{\ad}^+ L & = & 0 \non \\
\cD^+ \cD^+ L & =& \frac{2}{\raisebox{-1pt}{${\bar \cZ} - \cZ$}}
\Big( \cD^+ \cZ \cD^+ L
+ \cDB^+ {\bar \cZ} \cDB^+ L + \hf L \cD^+ \cD^+ \cZ \Big)\non \\
&& + \frac{{\rm i} \,g}{\raisebox{-1pt}{${\bar \cZ} - \cZ$}} \,{\rm tr}\,
\Big( ({\bar \cD}^+)^2 {\bar \cZ}^2 - (\cD^+)^2 \cZ^2 \Big)
\; .
\label{nothingnew}
\eea
which, however, are equivalent to the old ones \re{lvtcc}. Indeed, the
following redefinition
\be
\check{L} = L - {\rm i} g\,(\cZ - {\bar \cZ})
\ee
brings the constraints \re{nothingnew} to the form \re{lvtcc}.

\subsection{Linear vector-tensor multiplet with scale
and chiral\\ invariance} 
\label{scinv}

If the background $N=2$ vector multiplet in \re{lvtcs1} is abelian (here we
denote the corresponding superfield strength by $\cY$), then we are able to
couple it to the VT multiplet in a different fashion\footnote{The origin of
this alternative coupling is most easily understood in the six-dimensional
framework of Section \ref{6D}.}. Let us consider the constraints
(supersymmetry with rigid central charge)
\bea
\bD_\a^+ \bar{\bD}_{\ad}^+ L & = & 0 \non \\
\bD^+ \bD^+ L & = & - \frac{2}{\raisebox{-1pt}{$\cY + \bar{\cY}$}} \Big(
\bD^+ \cY \bD^+ L + \bar{\bD}^+ \bar{\cY} \bar{\bD}^+ L + \hf L \bD^+
\bD^+ \cY \Big)
\label{lsc1}
\eea
which satisfy all the superspace consistency conditions. In particular, it is
easy to check the rigid central charge form of \re{consistency1} using the
fact that $\bD^+\bD^+ L = \bar{\bD}^+\bar{\bD}^+ L$, as follows from
\re{lsc1}. Substituting the components \re{comp} of the VT multiplet and
defining the components of $\cY$ as in \re{extcomp}, we then arrive at the
component constraints
\be
\pa^m \big[ (W + \bar{W}) V_m - {\rm i} L \pa_m (W - \bar{W})
\big] = - \cF^{mn} \tilde{G}_{mn}\qquad \pa^m \tilde{G}_{mn} = 0
\ee
which can be easily solved.

If we gauge the central charge, the covariantized version of the constraints
\re{lsc1} compatible with all consistency conditions is as follows:
\bea
\cD_\a^+ \cDB_{\ad}^+ L & = & 0 \non \\
\cD^+ \cD^+ L & = & \frac{2 \bar{\cY}}{\raisebox{-1pt}{$\bar{\cZ} \cY -
\cZ \bar{\cY}$}} \Big( \cD^+ \cZ  \cD^+ L + \cDB^+ \bar{\cZ} \cDB^+ L
+ \hf L \cD^+ \cD^+ \cZ \Big) \non \\
&& - \frac{2 \bar{\cZ}}{\raisebox{-1pt}{$\bar{\cZ} \cY - \cZ \bar{\cY}$}}
\Big( \cD^+ \cY \cD^+ L + \cDB^+ \bar{\cY} \cDB^+ L + \hf L \cD^+ \cD^+
\cY \Big)\; .
\label{lsc2}
\eea
The corresponding Lagrangian density for this model is given by
\bea
\cL^{++} & = & - \frac{\rm i}{4} \Big( \cY \cD^+ L \cD^+ L - \bar{\cY}
        \cDB^+ L \cDB^+ L \Big) + \frac{\rm i}{8} \frac{\cY \bar{\cZ}
        + \cZ \bar{\cY}}{\raisebox{-1pt}{$\bar{\cZ} \cY - \cZ \bar{\cY}$}}\,
        L^2 \cD^+ \cD^+ \cY \non \\
&& - \frac{\rm i}{2} \frac{\cY \bar{\cY} L}{\raisebox{-1pt}{$\bar{\cZ}
        \cY - \cZ \bar{\cY}$}} \Big( \cD^+ \cZ \cD^+ L + \cDB^+ \bar{\cZ}
        \cDB^+ L + \hf L \cD^+ \cD^+ \cZ \Big) \non \\
&& + \frac{\rm i}{2} \frac{L}{\raisebox{-1pt}{$\bar{\cZ} \cY - \cZ
        \bar{\cY}$}}\, \Big( \cZ \bar{\cY}\, \cD^+ \cY \cD^+ L + \bar{\cZ}
        \cY\, \cDB^+ \bar{\cY} \cDB^+ L \Big)\; .
\label{lsc3}
\eea
As is seen from \re{lsc1} and \re{lsc2}, the constraints are well defined
only if $\cY$ has a non-vanishing vacuum expectation value. Freezing the
external vector multiplet by setting ${\cY} = 1$, constraints \re{lsc2}
reduce to \re{lvtcc} and Lagrangian density \re{lsc3} to that given by
eq.\ \re{lvtaction}.

A remarkable feature of the system presented here is that it respects the
invariance under global scale and chiral transformations of the superspace
coordinates
\bea
& \q' = {\rm e}^{-\o/2} \q \qquad \bar{\q}' = {\rm e}^{-\bar{\o}/2} \bar{\q}
\qquad x' = {\rm e}^{-(\o + \bar{\o})/2} x \non \\
&\cD'_{\a i} = {\rm e}^{\o/2} \cD_{\a i} \qquad \cDB'_{\ad i} =
{\rm e}^{\bar{\o}/2} \cDB_{\ad i} \qquad
\cD'_m = {\rm e}^{(\o + \bar{\o})/2} \cD_m \non\\
& \cZ'(z') = {\rm e}^{\bar{\o}} \cZ (z) \qquad
{\bar \cZ}'(z') = {\rm e}^{\o} {\bar \cZ}(z)
\label{sc1}
\eea
if we require $\cY$ and ${\bar {\cY}}$ to transform similarly to $\cZ$ and
$\bar{\cZ}$
\be
\cY'(z') = {\rm e}^{\bar{\o}} \cY(z) \qquad
\bar{\cY}'(z') = {\rm e}^{\o} \bar{\cY}(z)
\label{sc2}
\ee
and $L$ to have a vanishing scale and chiral weight
\be
L'(z') = L(z)\;.
\label{sc3}
\ee
This is in complete agreement with the results of \cite{claus2}.

Finally, we can couple the VT multiplet under consideration to an external
$N=2$ non-abelian vector multiplet $\cW$. The corresponding constraints read
\bea
\cD_\a^+ \cDB_{\ad}^+ L & = & 0 \non \\
\cD^+ \cD^+ L & = & \frac{2 \bar{\cY}}{\raisebox{-1pt}{$\bar{\cZ} \cY -
\cZ \bar{\cY}$}} \Big( \cD^+ \cZ  \cD^+ L + \cDB^+ \bar{\cZ} \cDB^+ L
+ \hf L \cD^+ \cD^+ \cZ \Big) \non \\
&& - \frac{2 \bar{\cZ}}{\raisebox{-1pt}{$\bar{\cZ} \cY - \cZ \bar{\cY}$}}
\Big( \cD^+ \cY \cD^+ L + \cDB^+ \bar{\cY} \cDB^+ L + \hf L \cD^+ \cD^+
\cY \Big) \non \\
&&+ {\rm i} g\, \frac{\bar{\cY}}{\raisebox{-1pt}{$
\bar{\cZ} \cY - \cZ \bar{\cY}$}}\, {\rm tr}\, \Big( \cDB^+ \cDB^+ \frac{
\bar{\cW}^2}{\bar{\cY}} - \cD^+ \cD^+ \frac{\cW^2}{\cY} \Big)\; .
\label{lsc4}
\eea
This system is also scale and chiral invariant if we require $\cW$ and $\bar
\cW$ to have standard vector multiplet transformation laws
\be
\cW'(z') = {\rm e}^{\bar{\o}} \cW(z) \qquad
\bar{\cW}'(z') = {\rm e}^{\o} \bar{\cW}(z)\;.
\label{sc4}
\ee
If we freeze the $\cY$--multiplet, by specifying ${\cY} = 1$, the constraints
\re{lsc4} reduce to \re{lvtcscc}.

We note that all the constraints in this subsection admit a simple derivation
starting from six dimensions (see Section \ref{6D}).

\sect{Models with nonlinear central charge\\
transformations} 
\label{nonlin}

In the present section we consider two VT multiplet models whose main
feature is the nonlinearity of the central charge transformations.

\subsection{Nonlinear vector-tensor multiplet} 

The nonlinear VT multiplet \cite{claus1} can be defined in harmonic
superspace by the constraints \cite{is}
\bea
{\bf D}_\a^+ {\bar {\bf D}}_{\ad}^+ L & = & 0 \non \\
{\bf D}^+ {\bf D}^+ L & = &
- \frac{1}{L}\Big( {\bf D}^+ L {\bf D}^+ L +  {\bar {\bf D}}^+ L
{\bar {\bf D}}^+ L \Big)\; .
\label{nvt1}
\eea
The superfield reparametrization
\be
L = \exp (- \k \,\tilde{L})
\label{sdur}
\ee
with $\k$ a real coupling constant, brings these constraints into the form
given in \cite{dk}
\bea
{\bf D}_\a^+ {\bar {\bf D}}_{\ad}^+ \,\tilde{L} & = &
\k \, {\bf D}_\a^+ \tilde{L} {\bar {\bf D}}_{\ad}^+ \tilde{L} \non \\
{\bf D}^+ {\bf D}^+ \tilde{L} & = &
\k \,{\bar {\bf D}}^+ \tilde{L} {\bar {\bf D}}^+ \tilde{L}
+2\k\, {\bf D}^+ \tilde{L} {\bf D}^+ \tilde{L}\;.
\label{reparam}
\eea
{}From eqs.\ \re{nvt1}, \re{reparam} it is obvious that in this case we are
dealing with a self-interacting VT multiplet.

To gauge the central charge transformations, one can again start from the
general Ansatz \re{ansatz} and look for consistent solutions possessing the
limit \re{nvt1} for $\cZ = -{\rm i}$. A complete analysis will be given
elsewhere. Here let us just present the constraints which underlie the
component construction of \cite{claus1} for the nonlinear VT multiplet with
gauged central charge:
\bea
&& \cD_\a^+ \cDB_{\ad}^+ L  = 0 \non \\
&& \cD^+ \cD^+ L  = - \frac{1}{\cZ}
\Big(\, 2 \cD^+ \cZ \cD^+ L + \hf L \cD^+
\cD^+ \cZ + \frac{\cZ}{L} \cD^+ L \cD^+ L - \frac{{\bar \cZ}}{L} \cDB^+ L
\cDB^+ L \Big)\; .
\label{nvt2}
\eea
The consistency conditions \re{consistency2} can be easily checked. The
central charge transformation of $\cD^+_\a L$ derived on the base of
\re{nvt2}
\bea
\cZ \D \cD_\a^+ L & = & {\rm i} \cD_{\a\ad} \cDB^{+\ad} L
- \cD_\a^+ \cZ\, \D L
+ \frac{1}{\raisebox{-1pt}{$4{\bar \cZ}$}} \cD_\a^-
\Big(\, 2 \cDB^+ {\bar \cZ} \cDB^+ L \non \\
&& + \hf L \cD^+ \cD^+ \cZ - \frac{\cZ}{L} \cD^+ L \cD^+ L +
\frac{{\bar \cZ}}{L} \cDB^+ L \cDB^+ L \Big)
\eea
passes the test \re{consistency1} too.

To discuss the differential constraints, we now turn to the component
fields of $L$. We choose them as in the linear case, with the exception
of $G_{mn}$:
\begin{equation} \label{nvtcomponent}
G_{\a\b} = -{\rm i} \Big( \frac{1}{4} \cZ [\cD_{\a i}\,,\,\cD_\b^i]
+ F_{\a\b} \Big) L| \qquad \bar{G}_{\ad\bd} = \overline{G_{\a\b}}\; .
\end{equation}
The constraints on $G_{mn}$ and $V_m$ (of which we again give only the
bosonic parts, the complete expressions can be found in \cite{claus1})
then read
\bea
 \nabla^n \tilde{G}_{mn}  & = & - \tilde{F}_{mn} V^n
\non\\
 \nabla^m H_{m}  & = & - \frac{1}{4} G^{mn} \tilde{G}_{mn}
- \frac{1}{4} (2 L \tilde{G}_{mn} - L^2 F_{mn}) \tilde{F}^{mn}
\eea
and the central charge transformations of $G_{mn}$ and $V_m$ are
\bea
\D G_{mn} & = & -2 \nabla_{[m} V_{n]} \non \\
\D H_m  & = & \tilde{G}_{mn} V^n - \frac{1}{4} \ve_{mnkl} \nabla^n (2 L
\tilde{G}^{kl} - L^2 F^{kl})\;.
\eea
Here we have used the notation
\begin{equation}
\nabla_m := \pa_m + A_m \,\D \qquad
H_m := |Z|^2 L V_m - \frac{\rm i}{2}  L^2 ({\bar Z} \pa_m Z
- Z \pa_m {\bar Z})\; .
\end{equation}
The constraints are solved by introducing a 1-form $T_m$ and a 2-form
$K_{mn}$ as in \cite{claus1}
\bea
  G_{mn} & = & 2 (\pa_{[m} T_{n]} - A_{[m} V_{n]}) \non \\
  H_m & = & \hf \ve_{mnkl} \big[ \pa^n K^{kl} - T^n \pa^k T^l - \hf A^n
        (2 L \tilde{G}^{kl} - L^2 F^{kl}) \big]\; .
\eea
These equations are still coupled, however. They can be separated by analogy
with \cite{bd,dt}, and we obtain
\bea
V_m & = & \frac{1}{E} \Big[ T_{mn} A^n + \frac{1}{2L} \ve_{mnkl} (\pa^n
  K^{kl} - T^n \pa^k T^l + L^2 A^n \pa^k A^l) \non\\
&& - \frac{1}{2|Z|^2 L} A_m \ve_{nklr}
        A^n (\pa^k K^{lr} - T^k \pa^l T^r) \non \\
&& + \frac{\rm i}{2}
L (\d_m^n - |Z|^{-2} A_m
        A^n)\, ({\bar Z} \pa_n Z - Z \pa_n {\bar Z}) \Big] \non \\
G_{mn} & = & T_{mn} - \frac{2}{E} A_{[m} \Big[ T_{n]k} A^k
+ \frac{\rm i}{2} L
        ({\bar Z} \pa_{n]} Z - Z \pa_{n]} {\bar Z}) \non\\
&&  + \frac{1}{2L} \ve_{n]klr} (\pa^k K^{lr} - T^k \pa^l T^r
           + L^2 A^k \pa^l A^r) \Big]
\eea
where $E = |Z|^2 - A^m A_m$, and $T_{mn} = \pa_m T_n - \pa_n T_m$ is the
field strength 2-form of $T_m$.

The central charge transformations of the potentials finally read (up to
gauge transformations)
\bea
  \D T_m & = & - V_m \non \\
  \D K_{mn} & = & T_{[m} V_{n]} - \hf (2 L \tilde{G}_{mn} - L^2
        F_{mn})\; .
\eea

For the constraints \re{nvt2} it is easy to find a supersymmetric Lagrangian
density
 \begin{equation}
  \cL^{++} = \frac{1}{4} L \big( \cZ \cD^+ L \cD^+ L
+ {\bar \cZ} \cDB^+ L \cDB^+ L -
        \frac{1}{6} L^2 \cD^+ \cD^+ \cZ \big)\; .
\label{nlvtaction}
\end{equation}
The action rule \re{action} then yields precisely the Lagrangian given in
\cite{claus1}:
\bea
\cL_{{\rm bos}} & = & L \Big[\, \hf |Z|^2 \big( \cD^m L \cD_m L - V^m
        V_m - |Z|^2 U^2 \big) - \frac{1}{12} L^2 Y_{ij} Y^{ij} \non \\
  && - \frac{1}{6} L^2 (Z \Box \bar Z + \bar Z
        \Box Z) + \frac{1}{4} (L F_{mn} - \tilde{G}_{mn})^2 - |Z|^2 U
        A^m \cD_m L \non \\
  && - A_m \cD_n L (L F^{mn} - \tilde{G}^{mn}) - A_m
        V_n (L \tilde{F}^{mn} + G^{mn}) \non \\
  && - \frac{1}{3} L^2 A_m \pa_n F^{mn} \Big]\; .
\eea

The dynamical system given by constraints \re{nvt2} and Lagrangian density
\re{nlvtaction} is invariant under global scale and chiral transformations
\re{sc1} and \re{sc3}.

There exist two more candidates for the r\^ole of Lagrangian density:
\bea
  \cL^{++}_1 & = & \frac{1}{4} \big( \cD^+ \cD^+ L + \cDB^+ \cDB^+ L \big)
\non \\
  \cL^{++}_2 & = & \frac{{\rm i}}{4}
\big( \cD^+ \cD^+ L - \cDB^+ \cDB^+ L \big)\; .
\label{triv}
\eea
However, the corresponding component Lagrangians are just total derivatives:
\bea
  \cL_1 & = & \pa_m \Big[\,
\frac{\bar Z - Z}{|Z|^2} \frac{{\rm i}}{2L}\, \ve^{mnkl}
        \big( \pa_n K_{kl} - T_n \pa_k T_l - L^2 A_n \pa_k A_l \big) \non \\
  && - \frac{Z + \bar Z}{|Z|^2} \big( \hf L
        \pa^m |Z|^2 - L F^{mn} A_n + \tilde{T}^{mn} A_n \big) \Big] \non\\
       & \equiv & \pa_m M^m (Z, \bar Z, \cdots)\non\\
 \cL_2 & = & \pa_m M^m (-{\rm i} Z, {\rm i} \bar Z, \cdots)\; .
\eea

\subsection{Nonlinear vector-tensor multiplet
with Chern-Simons\\ couplings} 

The nonlinear VT multiplet coupled to an external $N=2$ super Yang-Mills
multiplet is described by the constraints
\bea
{\bf D}_\a^+ {\bar {\bf D}}_{\ad}^+ L & = & 0 \non \\
{\bf D}^+ {\bf D}^+ L & = &
- \frac{1}{L}\Big( {\bf D}^+ L {\bf D}^+ L +  {\bar {\bf D}}^+ L
{\bar {\bf D}}^+ L \Big)\non\\
&& - \frac{{\bf g}}{2L} \,{\rm tr}\,
\Big( ({\bar \cD}^+)^2 {\bar \cW}^2 + (\cD^+)^2 \cW^2 \Big)
\label{nvtcs}
\eea
with $\bf g$ a real coupling constant. If one implements the superfield
redefinition \re{sdur}, the constraints turn into
\bea
{\bf D}_\a^+ {\bar {\bf D}}_{\ad}^+ \,\tilde{L} & = &
\k \, {\bf D}_\a^+ \tilde{L} {\bar {\bf D}}_{\ad}^+ \tilde{L} \non \\
{\bf D}^+ {\bf D}^+ \tilde{L} & = &
\k \,{\bar {\bf D}}^+ \tilde{L} {\bar {\bf D}}^+ \tilde{L}
+2\k\, {\bf D}^+ \tilde{L} {\bf D}^+ \tilde{L} \non \\
&& + \frac{{\bf g}}{2\k}\, {\rm e}^{2\k \,\tilde{L}}\;{\rm tr}\,
\Big( ({\bar \cD}^+)^2 {\bar \cW}^2 + (\cD^+)^2 \cW^2 \Big) \;.
\label{reparam2}
\eea
Rescaling here ${\bf g} = \k\,g$ and considering the limit of vanishing
self-coupling, $\k \rightarrow 0$, the latter constraints reduce to
\bea
{\bf D}_\a^+ {\bar {\bf D}}_{\ad}^+ \,\tilde{L} & = & 0 \non \\
{\bf D}^+ {\bf D}^+ \tilde{L} & = & \frac{g}{2}\, {\rm tr}\,
\Big( ({\bar \cD}^+)^2 {\bar \cW}^2 + (\cD^+)^2 \cW^2 \Big) \;.
\label{reparam3}
\eea
These constraints take the form given by eq.\ \re{standard} if we redefine
\be
\tilde{L} \equiv {\Bbb L} + \hf \, g\;{\rm tr}\, (\cW + {\bar \cW})^2 \;.
\ee

A consistent deformation of the constraints \re{nvtcs}, which corresponds to
the gauged central charge, reads
\bea
\cD_\a^+ \cDB_{\ad}^+ L  & = & 0 \non \\
\cD^+ \cD^+ L  & = & - \frac{1}{\cZ}
\Big(\, 2 \cD^+ \cZ \cD^+ L + \hf L \cD^+
\cD^+ \cZ + \frac{\cZ}{L} \cD^+ L \cD^+ L - \frac{{\bar \cZ}}{L} \cDB^+ L
\cDB^+ L \Big) \non\\
&& - \frac{1}{\cZ}\, \frac{{\bf g}}{2L} \,{\rm tr}\,
\Big( ({\bar \cD}^+)^2 \frac{{\bar \cW}^2}{\bar{\cZ}} - (\cD^+)^2
\frac{\cW^2}{\cZ} \Big)
\; .
\label{nvtcs2}
\eea
All consistency conditions are identically satisfied, as can be easily
checked.

The constraint on $G_{mn}$ and its central charge transformation do not
change, while the equations for $H_m$ now read
\bea
\D H_m & = & \tilde{G}_{mn} V^n - \frac{1}{4} \ve_{mnkl} \nabla^n (2 L
\tilde{G}^{kl} - L^2 F^{kl}) \non \\
&& - {\bf g}\, \pa^n\, {\rm tr}\, \Big[\, {\rm i} \Big(\frac{W^2}{Z^2} -
\frac{\bar{W}^2}{\bar{Z}^2} \Big) F_{mn} + \Big(\frac{W^2}{Z^2} +
\frac{\bar{W}^2}{\bar{Z}^2} \Big) \tilde{F}_{mn} - 2 {\rm i}\, \Big(
\frac{W}{Z} - \frac{\bar{W}}{\bar{Z}} \Big) \cF_{mn} \non \\
&& + 2 \Big( \frac{W}{Z} + \frac{\bar{W}}{\bar{Z}} \Big) \tilde{\cF}_{mn}
\Big] \non \\[6pt]
\nabla^m H_m & = & - \frac{1}{4} G^{mn} \tilde{G}_{mn} - \frac{1}{4} (2 L
\tilde{G}_{mn} - L^2 F_{mn}) \tilde{F}^{mn} \non \\
&& - \hf {\bf g}\, F^{mn} {\rm tr}\, \Big[\, {\rm i} \Big( \frac{W^2}{Z^2}
- \frac{\bar{W}^2}{\bar{Z}^2} \Big) F_{mn} + \Big( \frac{W^2}{Z^2} +
\frac{\bar{W}^2}{\bar{Z}^2} \Big) \tilde{F}_{mn} - 2 {\rm i}\, \Big(
\frac{W}{Z} - \frac{\bar{W}}{\bar{Z}} \Big) \cF_{mn} \non \\
&& - 2 \Big( \frac{W}{Z} + \frac{\bar{W}}{\bar{Z}} \Big) \tilde{\cF}_{mn}
\Big] - {\rm i} {\bf g}\, \pa^m {\rm tr}\, \Big[ W \cD_m \bar{W} - \bar{W}
\cD_m W - \frac{W^2}{Z} \pa_m \bar{Z} + \frac{\bar{W}^2}{\bar{Z}} \pa_m Z
\non \\
&& + \hf \cD_m \Big( \frac{\bar{Z}}{Z} W^2 - \frac{Z}{\bar{Z}} \bar{W}^2
\Big) \Big] - {\bf g}\, {\rm tr}\, (\cF^{mn} \tilde{\cF}_{mn})\; ,
\eea
the solution of which is
\bea
H_m & = & \hf \ve_{mnkl} \Big[ \pa^n K^{kl} - T^n \pa^k T^l - \hf A^n
(2 L \tilde{G}^{kl} - L^2 F^{kl}) - 4 {\bf g}\, {\rm tr}\, (\cA^n \pa^k
\cA^l + \frac{2}{3} \cA^n \cA^k \cA^l) \Big] \non \\
&& - {\bf g}\, {\rm tr}\, \Big[ \, {\rm i} \Big(\frac{W^2}{Z^2} -
\frac{\bar{W}^2}{\bar{Z}^2} \Big) F_{mn} A^n + \Big(\frac{W^2}{Z^2} +
\frac{\bar{W}^2}{\bar{Z}^2} \Big) \tilde{F}_{mn} A^n - 2 {\rm i}\, \Big(
\frac{W}{Z} - \frac{\bar{W}}{\bar{Z}} \Big) \cF_{mn} A^n \non \\
&& - 2 \Big( \frac{W}{Z} + \frac{\bar{W}}{\bar{Z}} \Big) \tilde{\cF}_{mn}
A^n + {\rm i} W \cD_m \bar{W} - {\rm i} \bar{W} \cD_m W - {\rm i}
\frac{W^2}{Z} \pa_m \bar{Z} + {\rm i} \frac{\bar{W}^2}{\bar{Z}} \pa_m Z
\non \\
&& + \frac{\rm i}{2} \cD_m \Big( \frac{\bar{Z}}{Z} W^2 - \frac{Z}{\bar{Z}}
\bar{W}^2 \Big) \Big]\; .
\eea

The corresponding Lagrangian density is
\bea
\cL^{++} & = & \frac{1}{4} L \Big( \cZ \cD^+ L \cD^+ L
+ {\bar \cZ} \cDB^+ L \cDB^+ L -
\frac{1}{6} L^2 \cD^+ \cD^+ \cZ \Big) \\
&& + \frac{1}{8}\,{\bf g}\; {\rm tr}\, \Big(
L(\cD^+)^2 \frac{\cW}{\cZ}^2
+ L ({\bar \cD}^+)^2 \frac{{\bar \cW}}{{\bar \cZ}}^2
-2 (\cD^+)^2 (L \frac{\cW}{\cZ}^2)
-2 ({\bar \cD}^+)^2 (L \frac{{\bar \cW}}{{\bar \cZ}}^2) \Big)
\; .\non
\eea
One can readily check that $\cL^{++}$ satisfies the basic requirements
\re{analyticity1} and \re{holomorphy1}. The dynamical system under
consideration is invariant under global scale and chiral transformations
\re{sc1}, \re{sc3} and \re{sc4}.

In principle, one could consider more general constraints than those
given in eq.\ \re{nvtcs2}
\bea
\cD_\a^+ \cDB_{\ad}^+ L  & = & 0 \non \\
\cD^+ \cD^+ L  &=& - \frac{1}{\cZ}
\Big(\, 2 \cD^+ \cZ \cD^+ L + \hf L \cD^+
\cD^+ \cZ + \frac{\cZ}{L} \cD^+ L \cD^+ L - \frac{{\bar \cZ}}{L} \cDB^+ L
\cDB^+ L \Big) \non \\
&& + \frac{1}{\cZ}\, \frac{1}{2L} \,{\rm tr}\,
\Big(  (\cD^+)^2 \cF(\cW, \cZ)
- ({\bar \cD}^+)^2 {\bar \cF}({\bar \cW}, {\bar \cZ}) \Big)
\eea
with $\cF(\cW, \cZ)$ a gauge invariant holomorphic function. It is obvious that
all consistency conditions are identically satisfied. Like in the case of eq.
\re{nolabel}, it is not possible, in general, to solve the constrained vector
and antisymmetric tensor component fields of $L$ in terms of gauge two- and
one-forms, respectively (see, however, footnote 2).

If we identify $\cW$ in the constraints \re{nvtcs2} with $\cZ$, then we
obtain exactly in the nonlinear VT multiplet constraints \re{nvt2}, as a
consequence of the Bianchi identity \re{bianchi}.

\sect{The linear vector-tensor multiplet from six-\\ dimensional
harmonic superspace} \label{6D} 

\subsection{Supersymmetry in six dimensions} 

\subsubsection{Six-dimensional superspace}

In this section we follow a number of standard references (see, e.g.,
\cite{HST,HSW,zupnik}). There are two types of spinor indices
${\hat\a}=1,2,3,4$ of the group USp(4): `left-handed' $\j^{\hat\a}$ and
`right-handed' $\j_{\hat\a}$. Correspondingly, there are two types of gamma
matrices, $\G^\hm_{{\hat\a}{\hat\b}}$ and $\G^{\hm{\hat\a}{\hat\b}}$
($\hm=0,1,2,3,5,6$), which are related by raising and lowering the pair of
antisymmetric spinor indices ${\hat\a}{\hat\b}$:
\begin{equation}\label{gammat}
  \G^{\hm{\hat\a}{\hat\b}} =
{1\over 2}\ve^{{\hat\a}{\hat\b}{\hat\g}{\hat\d}}\G^\hm_{{\hat\g}{\hat\d}}\;,
\qquad \G^\hm_{{\hat\a}{\hat\b}} =  {1\over
  2}\ve_{{\hat\a}{\hat\b}{\hat\g}{\hat\d}}\G^{\hm{\hat\g}{\hat\d}}\;.
\end{equation}
The epsilon symbol is defined by $\ve^{1234}=\ve_{1234}=1$ and
it satisfies the identity
\begin{equation}
  \ve^{{\hat\a}{\hat\b}{\hat\g}{\hat\d}} \ve_{{\hat\a}{\hat\b}{\hat\r}
  {\hat\s}} = 4{\d}^{[{\hat\g}}_{\hat\r} {\d}^{{\hat\d}]}_{\hat\s}\;.
\end{equation}
The gamma matrices satisfy the usual algebra
\begin{equation}\label{gamalg}
  \G^\hm_{{\hat\a}{\hat\b}}\G^{\hn{\hat\b}{\hat\g}} +
  \G^\hn_{{\hat\a}{\hat\b}}\G^{\hm{\hat\b}{\hat\g}} =
2\eta^{\hm\hn}{\d}^{\hat\g}_{\hat\a}
\end{equation}
and have the property
\begin{equation}\label{gamprop1}
  \G^\hm_{{\hat\a}{\hat\b}}\G^{{\hat\g}{\hat\d}}_\hm =
  -4{\d}^{[{\hat\g}}_{\hat\a} {\d}^{{\hat\d}]}_{\hat\b}\;.
\end{equation}
Two useful corollaries of (\ref{gamalg}) and (\ref{gamprop1}) are
\begin{equation}\label{gamprop2}
  \G_{\hm{\hat\a}{\hat\b}}\G^{\hn{\hat\a}{\hat\b}}= -4{\d}^\hn_\hm\;, \qquad
  \G_{\hm{\hat\a}{\hat\b}}\G^\hm_{{\hat\g}{\hat\d}} = -2\ve_{{\hat\a}{\hat\b}
  {\hat\g}{\hat\d}}\;.
\end{equation}

The Grassmann superspace coordinate $\q^{{\hat\a}}_i$ has a left-handed
spinor index of USp(4) and an SU(2) index $i$. The corresponding spinor
derivative $\pa_{{\hat\a}}^i = \pa/\pa\q^{{\hat\a}}_i$ has a right-handed
spinor index of USp(4). Besides the Weyl condition (which is taken care of
by the above notation), these spinors satisfy a pseudo-Majorana reality
condition (see \cite{HST}).

The algebra of the spinor covariant derivatives is
\begin{equation}\label{algder}
  \{D_{{\hat\a}}^i, D_{{\hat\b}}^j\} =
  2{\rm i}\, \ve^{ij} \G^\hm_{{\hat\a}{\hat\b}} \pa_\hm \equiv
  2{\rm i}\, \ve^{ij} \pa_{{\hat\a}{\hat\b}}\;.
\end{equation}
When the harmonic variables $u^{\pm i}$ are introduced, $\q^{{\hat\a}}_i$
and $D_{{\hat\a}}^i$ split up into two U(1)-charged projections:
\begin{equation}
  \q^{\pm{\hat\a}}= \q^{{\hat\a} i}u^\pm_i\;,
\qquad D^\pm_{\hat\a} = u^{\pm}_iD_{{\hat\a}}^i\;,
\end{equation}
so that
\begin{equation}
  D^+_{\hat\a} \q^{-{\hat\b}} = {\d}^{\hat\b}_{\hat\a}\;.
\end{equation}
Thus, the algebra \re{algder} becomes
\begin{eqnarray}
  && \{D^+_{\hat\a},D^+_{\hat\b}\} =
     \{D^-_{\hat\a},D^-_{\hat\b}\} = 0\;, \label{integ}  \\
  && \{D^+_{\hat\a},D^-_{\hat\b}\} =
     -2{\rm i}\, \pa_{{\hat\a}{\hat\b}}\;. \label{+-}
\end{eqnarray}

\subsubsection{Analytic basis and sechsbein}

The integrability condition \re{integ} allows one to choose an analytic
basis in which the spinor derivative $D^+_{\hat\a}$ becomes `short',
\begin{equation}
  D^+_{\hat\a} = \pa^+_{\hat\a}\;.
\end{equation}
In this basis the harmonic derivatives acquire vielbeins:
\begin{equation}
  D^{\pm\pm} = \pa^{\pm\pm} + H^{\pm\pm \hm}\pa_\hm
\end{equation}
which satisfy the harmonic constraint
\begin{equation}\label{harmcon}
  D^{++}H^{--\hm} = D^{--}H^{++\hm}
\end{equation}
as a consequence of the commutation relation $[D^{++},D^{--}]= D^0$.
As long as the superspace remains flat, we have just
\begin{equation}
  H^{\pm\pm \hm}|_0 = -{\rm i}\q^\pm\G^\hm\q^\pm\;,
\end{equation}
but these vielbeins will become non-trivial later on, when we perform the
reduction to four dimensions and gauge the two translations along
the fifth and sixth directions. The minus projection of the spinor
derivative is obtained by
\begin{equation}
  D^-_{\hat\a} = [D^{--},D^+_{\hat\a}] =
-\pa^-_{\hat\a} - D^+_{\hat\a} H^{--\hm}\pa_\hm
\end{equation}
and then \re{+-} gives
\begin{equation}
  \{D^+_{\hat\a}, D^-_{\hat\b}\} =
  -D^+_{\hat\a} D^+_{\hat\b} H^{--\hm}\pa_\hm
  \equiv -2{\rm i} e^\hm_{{\hat\a}{\hat\b}}\pa_\hm
  \equiv -2{\rm i} D_{{\hat\a}{\hat\b}}\;.
\end{equation}
The sechsbein
\begin{equation}\label{6bein}
  e^\hm_{{\hat\a}{\hat\b}} = -e^\hm_{{\hat\b}{\hat\a}} =
-{{\rm i}\over 2}D^+_{\hat\a} D^+_{\hat\b} H^{--\hm}
\end{equation}
just coincides with the gamma matrices in the flat case,
\begin{equation}
  e^\hm_{{\hat\a}{\hat\b}}|_0 = \G^\hm_{{\hat\a}{\hat\b}}\;,
\end{equation}
but it will play a key r\^ole in the context of gauged central charges. Note
that the properties \re{gamprop1}, \re{gamprop2} of the gamma matrices can
serve as a formal definition of the inverse sechsbein:
\begin{equation}\label{inv6bein}
  e^\hm_{{\hat\a}{\hat\b}} e_\hm^{{\hat\g}{\hat\d}} =
  -4{\d}^{[{\hat\g}}_{\hat\a} {\d}^{{\hat\d}]}_{\hat\b}\;,
  \qquad e_\hm^{{\hat\a}{\hat\b}}e^\hn_{{\hat\a}{\hat\b}}
  = -4{\d}^\hn_\hm\;.
\end{equation}

\subsection{Dimensional reduction to four dimensions
and the origin of central charge}

The dimensional reduction from six to four dimensions is obtained by
replacing the four-component spinor index $\hat\a$ of USp(4) by a pair
of undotted and dotted spinor indices of SL($2,{\Bbb C}$),
$\hat\a = (\a,\ad)$. The gamma matrices $\G^\hm$ then split into:
\begin{eqnarray}
   &&\G^m_{\a\bd} = \s^m_{\a\bd}\;, \quad
   \G^m_{\a\b} = \G^m_{\ad\bd}=0  \ (m=0,1,2,3)\;;  \nonumber \\
    &&\G^5_{\a\bd} = 0\;, \quad \G^5_{\a\b} = -\ve_{\a\b}\;,
    \quad \G^5_{\ad\bd}
    = -\ve_{\ad\bd}\;;  \nonumber \\
   &&\G^6_{\a\bd} = 0\;, \quad \G^6_{\a\b} = {\rm i}\ve_{\a\b}\;,
   \quad \G^6_{\ad\bd}
    = -{\rm i}\ve_{\ad\bd}\;.
\end{eqnarray}
The four-index epsilon symbol splits into a product of two two-index ones:
\begin{equation}
  \ve_{\hat\a\hat\b\hat\g\hat\d} \ \rightarrow \ \ve_{\a\b}\ve_{\dot\g\dot\d}
\end{equation}
(the other components varnish or are obtained by permutations).

Accordingly, the spinor derivatives algebra \re{algder} becomes
\begin{eqnarray}
   &&\{D_{\a i},\bar D_{\ad j}\} =
  -2{\rm i}\ve_{ij}\pa_{\a\ad}\;,  \label{4dim1} \\
    &&\{D_{\a i},D_{\b j}\} =
  2\ve_{ij}\ve_{\a\b}({\rm i}\pa_5 + \pa_6)\;,  \label{4dim2} \\
&&\{\bar D_{\ad i},\bar D_{\bd j}\} = -2\ve_{ij}\ve_{\ad\bd}(-{\rm i}\pa_5 +
\pa_6)\;.\label{4dim3}
\end{eqnarray}
The essential point is the appearance of the derivatives $\pa_{5,6}$ in the
anticommutators \re{4dim2} and \re{4dim3}. Comparing this with eqs.\
\re{cda}, one concludes that $\pa_{5,6}$ are the generators of {\it two real
central charges} (or one complex one). Thus, the dimensional reduction from
six to four dimensions gives a natural explanation of the origin of the
central charges in the $N=2$ $D=4$ supersymmetry algebra.

Let us now look at the harmonic vielbeins $H^{\pm\pm\hm}$. In the flat case
they are simply
\begin{equation}\label{flatH}
  H^{\pm\pm m}|_0=-2{\rm i}\q^\pm\s^m\bar\q^\pm\;,
\quad H^{\pm\pm 5}|_0 = {\rm i}(\q^\pm\q^\pm -
  \bar\q^\pm\bar\q^\pm)\;,
\quad H^{\pm\pm 6}|_0 = \q^\pm\q^\pm + \bar\q^\pm\bar\q^\pm\;.
\end{equation}
However, if any of the central charges (or both) is gauged, the corresponding
vielbein becomes a non-trivial harmonic superfield, $H^{\pm\pm 5,6}(x^\hm,\q,
\bar\q,u)$ and the harmonic derivatives $D^{\pm\pm}$ become covariant,
\begin{equation}
  \cD^{\pm\pm} = \pa^{\pm\pm} + H^{\pm\pm 5,6}\pa_{5,6}\;.
\end{equation}
It is crucial that the newly introduced gauge superfields $H^{\pm\pm 5,6}$
are not allowed to depend on the central charge coordinates $x^{5,6}$:
\begin{equation}
  \pa_{5,6}H^{\pm\pm 5,6} = 0\;.
\end{equation}
In this way they give rise to two ordinary (i.e., central-charge independent)
{\it abelian} gauge supermultiplets. In addition, the vielbeins $H^{++ 5,6}$
have to be {\it analytic} superfields:
\begin{equation}
  [ D^+_\a, \cD^{++}] = 0 \quad \Rightarrow \quad D^+_\a H^{++ 5,6} = 0\;.
\end{equation}
This choice allows the spinor derivative $D^+_\a$ to remain flat,
\begin{equation}
  \cD^+_\a = D^+_\a \;.
\end{equation}
These analytic superfields are the prepotentials of the two abelian gauge
multiplets. In fact, one of them, $H^{++5}$, coincides with the central
charge connection $\cV^{++}$ introduced in Section \ref{fund}. The
(non-analytic) vielbeins $H^{-- 5,6}$ are obtained, as usual, by solving
the relation \re{harmcon} which is the same as \re{prisvoim}.

The prepotentials are not gauge invariant objects. To find out which are the
invariant ones, let us inspect the various components of the sechsbein. The
vielbein $H^{--m}$ has remained flat, and so have the sechsbein components
$e^m_{\a\bd}$:
\begin{equation}
  e^m_{\a\bd} = \s^m_{\a\bd}\;.
\end{equation}
However,
\begin{equation}
  e^5_{\a\b} = -{{\rm i}\over 2} D^+_\a D^+_\b H^{--5}
= -{{\rm i}\over 4} \ve_{\a\b}D^+D^+
  H^{--5} \equiv {\rm i}\ve_{\a\b} \bar\cZ\;.
\end{equation}
It is easy to see that the quantity
\begin{equation}\label{defZ}
   \bar\cZ = -{1\over 4}D^+D^+H^{--5}
\end{equation}
is gauge invariant (it is the same as in eq.\ \re{strV}, after the
identification $H^{--5}=\cV^{--}$). Indeed, in the context of dimensional
reduction the abelian gauge transformations are generated by local shifts
of the central-charge coordinates:
\begin{equation}
  \d x^{5,6} = \l^{5,6}(x^m,\q,\bar\q,u)\;.
\end{equation}
The abelian character of these transformations is expressed in the fact that
the parameters $\l^{5,6}$ do not depend on $x^{5,6}$:
\begin{equation}\label{nondep}
   \pa_{5,6}\l^{5,6} = 0\;.
\end{equation}
In addition, they must be {\it analytic} in order to preserve the `short'
form of the spinor derivatives $D^+_{\a,\ad}$:
\begin{equation}\label{analparam}
  D^+_{\a,\ad}\l^{5,6}=0\;.
\end{equation}
Now, the vielbeins $H^{\pm\pm 5,6}$ have the typical transformation law:
\begin{equation}
  \d H^{\pm\pm 5,6} = \cD^{\pm\pm}\l^{5,6} = D^{\pm\pm}\l^{5,6}
\end{equation}
by virtue of \re{nondep}. Then from \re{defZ}, \re{analparam} and from the
spinor derivatives algebra one obtains
\begin{equation}\label{invarZ}
  \d\bar\cZ = -{1\over 4}D^+D^+D^{--}\l^5 = {1\over 4}D^{+\a}D^-_\a\l^5 = 0\;.
\end{equation}
Thus, $\bar\cZ$ is the {\it field strength} of the abelian gauge multiplet.
We must stress that it does not depend on the central charge coordinates:
\begin{equation}
  \pa_{5,6}\bar\cZ=0\;,
\end{equation}
as expected from an ordinary abelian gauge supermultiplet. Further, it
satisfies the harmonic constraint
\begin{equation}\label{harmindZ}
  \cD^{++}\bar\cZ = D^{++}\bar\cZ = 0
\end{equation}
(the proof is similar to \re{invarZ} and makes use of the harmonic relation
\re{harmcon}). In addition, $\bar\cZ$ has all the properties \re{bianchi}
and (\ref{Ban}) derived in Section \ref{fund}. Note that the flat limit of
$\bar\cZ$ (i.e., with $H^{--}$ taken from (\ref{flatH})) is
\begin{equation}
  \bar\cZ|_0 = {\rm i}\;.
\end{equation}

Similarly, the other non-trivial sechsbein component $e^6_{\a\b}$ gives
rise to the second abelian field strength:
\begin{equation}
  e^6_{\a\b} = {\rm i}\ve_{\a\b}\bar\cY\;,
\qquad \bar\cY = -{1\over 4}D^+D^+H^{--6}
\end{equation}
with the same properties as $\bar\cZ$, except for the flat limit
\begin{equation}
 \bar\cY|_0 = 1\;.
\end{equation}
As we shall see below, this field strength is the one which has been already
introduced in subsection \ref{scinv}.

The conclusion of the above discussion is that gauging the central charges
amounts to restricted diffeomorphisms in the six-dimensional superspace (only
along the 5th and 6th dimensions; no dependence on $x^{5,6}$ is allowed). The
corresponding components of the sechsbein are identified with the abelian
field strengths.

\subsection{Torsion in six dimensions} \label{subsec} 

The procedure of obtaining the field strengths presented above is clearly not
covariant in six dimensions. There the gauge (i.e.\ diffeomorphism invariant)
object should be looked for one level higher in the covariant derivatives
algebra. To keep six-dimensional covariance manifest, we shall treat all the
prepotentials $H^{--\hm}$ as non-trivial, i.e. no distinction will be made
here between $H^{--m}$ and $H^{--5,6}$. Correspondingly, we shall consider
{\it analytic diffeomorphisms} of all the six coordinates $x^\hm$:
\begin{equation}\label{6diff}
  \d x^\hm = \l^\hm(x^\hn,\q, u)\;, \qquad D^+_{\hat\a}\l^\hm = 0\;.
\end{equation}
This amounts to having the full $6\times 6$ sechsbein matrix \re{6bein},
as well as its inverse defined by \re{inv6bein}. It should be stressed
that even within this enlarged six-dimensional context the spinor derivatives
$D^+_{\hat\a}$ remain flat:
\begin{equation}
  \cD^+_{\hat\a} = D^+_{\hat\a}\;.
\end{equation}
This fact expresses the fundamental postulate of preservation of Grassmann
analyticity.

Now, consider the commutator
\begin{eqnarray}
  [\cD^+_{\hat\a},\cD_{{\hat\b}{\hat\g}}] &=&
  -{{\rm i}\over 2}D^+_{\hat\a} D^+_{\hat\b} D^+_{\hat\g} H^{--\hm}\pa_\hm
\nonumber \\
    &=& -{{\rm i}\over 2}\ve_{{\hat\a}{\hat\b}{\hat\g}{\hat\d}}
(D^+)^{3{\hat\d}} H^{--\hm}\pa_\hm  \nonumber \\
    &=& {{\rm i}\over 8}\ve_{{\hat\a}{\hat\b}{\hat\g}{\hat\d}}
(D^+)^{3{\hat\d}} H^{--\hm}e_{\hm{\hat\r}{\hat\s}}\cD^{{\hat\r}{\hat\s}}
\nonumber \\
    &=& {1\over 2}\ve_{{\hat\a}{\hat\b}{\hat\g}{\hat\d}} T^{+{\hat\d}}
    {}_{{\hat\r}{\hat\s}}\cD^{{\hat\r}{\hat\s}}\;.\label{comtors}
\end{eqnarray}
Here $(D^+)^{3{\hat\d}}\equiv -{1\over 6}\ve^{{\hat\d}{\hat\a}{\hat\b}
{\hat\g}}D^+_{\hat\a} D^+_{\hat\b} D^+_{\hat\g}$ and the torsion tensor
$T^+$ is given by
\begin{equation}\label{tors}
  T^{+{\hat\a}}{}_{{\hat\b}{\hat\g}} =
  {{\rm i}\over 4}(D^+)^{3{\hat\a}} H^{--\hm}e_{\hm{\hat\b}{\hat\g}}\;.
\end{equation}
By construction, this quantity is invariant under the diffeomorphisms
\re{6diff}. This can also be verified directly from the expression
\re{tors} using the algebra of the covariant derivatives. The presence of
torsion reflects the fact that the superspace is not completely flat (recall
the abelian curvature related to the gauged central charge in the
four-dimensional context). Note, however, that it is not our aim here to
develop a full-fledged supergravity formalism in six dimensions (that would
require to consider the full {\it superdiffeomorphism} group and to introduce
a complete set of supervielbeins; for details see \cite{Sok}).

By raising the pair of indices ${\hat\b}{\hat\g}$, eq.\ \re{comtors} can be
rewritten in the equivalent form
\begin{equation}\label{eqform}
  [ D^+_{\hat\a},\cD^{{\hat\b}{\hat\g}}] = \d^{[{\hat\b}}_{\hat\a}
  T^{+{\hat\g}]}{}_{{\hat\r}{\hat\s}} \cD^{{\hat\r}{\hat\s}}\;.
\end{equation}
Then, anticommuting the left-hand side of eq.\ \re{eqform} with another
spinor derivative $D^+_{\hat\k}$, one derives the Bianchi identity
\begin{equation}\label{traceless}
  D^+_{({\hat\a}} T^{+{\hat\b})}{}_{{\hat\r}{\hat\s}} -
  T^{+({\hat\b}}{}_{{\hat\a}){\hat\g}} T^{+{\hat\g}}{}_{{\hat\r}{\hat\s}}
  = 0\;,
\end{equation}
where ${}_{({\hat\a}}{}^{{\hat\b})}$ denotes the traceless part.

Another Bianchi identity is obtained by noting that
\begin{equation}
  [\cD^{++},\cD_{{\hat\a}{\hat\b}}] =
{{\rm i}\over 2}[\cD^{++},\{ D^+_{\hat\a},\cD^-_{\hat\b}\}] =
{{\rm i}\over 2}
\{ D^+_{\hat\a}, D^+_{\hat\b}\} = 0\;.
\end{equation}
Applying this to eq.\ \re{eqform}, one finds
\begin{equation}
  \cD^{++} T^{+{\hat\a}}{}_{{\hat\b}{\hat\g}} = 0 \;.
\end{equation}

\subsection{The linear vector-tensor multiplet} 

\subsubsection{Six-dimensional formulation}

The six-dimensional analog of the linear VT multiplet is the so-called
self-dual tensor multiplet \cite{BPSez}. In harmonic superspace it is
described by a real superfield $L$ satisfying the constraint \cite{Sok}
\begin{equation}\label{lincon}
  D^+_{\hat\a} D^+_{\hat\b} L = 0\;.
\end{equation}
This constraint simply means that $L$ is a linear function of the Grassmann
variables $\q^-$:
\begin{equation}\label{solself}
  L = l(x,\q^+,u) + \q^{-{\hat\a}}\psi^+_{\hat\a}(x,\q^+,u)\;.
\end{equation}
Note that the coefficient functions are analytic, but one of them ($l$) is
not a superfield, since it transforms into $\psi^+$ under supersymmetry. One
can also present the solution \re{solself} in the superfield form
\begin{equation}
  L = (D^+)^{3\hat\a}\Psi^{-3}_{\hat\a}
\end{equation}
where $\Psi^{-3}_{\hat\a}$ is an arbitrary prepotential.

In addition to \re{lincon}, $L$ satisfies the harmonic constraint
\begin{equation}\label{harmlincon}
  D^{++}L=0
\end{equation}
which reduces its harmonic dependence to a polynomial one and thus produces a
finite supermultiplet. The two constraints \re{lincon} and \re{harmlincon}
are clearly compatible. The harmonic constraint can also be written down in
the equivalent form (see the discussion after eq.\ \re{harmVTc2})
\begin{equation}\label{--lincon}
  D^{--}L=0\;.
\end{equation}
Acting on \re{--lincon} with one, two, three or four spinor derivatives
$D^+_{\hat\a}$ and using \re{lincon}, one obtains the components of the
self-dual tensor multiplet in six dimensions. Those are a scalar $\o$, a
self-dual antisymmetric tensor (in spinor notation $F_{\hat\a\hat\b} =
F_{\hat\b\hat\a}$) and a spinor $\l_{\hat\a i}$ satisfying the on-shell
equations
\begin{equation}\label{eqmot}
  \Box\o = 0\;, \quad \pa^{(\hat\a\hat\b}F_{\hat\b\hat\g)}=0\;,
  \quad \pa^{\hat\a\hat\b}\l_{\hat\b i} = 0\;.
\end{equation}
In addition, all these fields are harmonic-independent. Taking all this into
account, one reduces the expansion of the superfield $L$ to
\begin{equation}
  L = \o(x) + \q^{-\hat\a}\q^{+\hat\b}[F_{\hat\a\hat\b}(x) +
  {\rm i}\pa_{\hat\a\hat\b}\o(x)] + \q^{+\hat\a}u^-_i\l^i_{\hat\a}(x)
   - \q^{-\hat\a}u^+_i\l^i_{\hat\a}(x)\;.
\end{equation}

Now we are going to covariantize the above picture with respect to the
diffeomorphism group \re{6diff}. While the spinor derivative constraint
\re{lincon} needs no covariantization, the harmonic ones become covariant:
\begin{equation}\label{covharmlincon}
  \cD^{++}L=0
\end{equation}
and
\begin{equation}\label{cov--lincon}
  \cD^{--}L=0\;.
\end{equation}
This immediately raises the issue of compatibility between \re{lincon} and
\re{cov--lincon}, just as in subsection \ref{linvt}, where the same problem
arose in the four-dimensional context. Indeed, one should have
\begin{equation}\label{compc}
  \ve^{{\hat\a}{\hat\b}{\hat\g}{\hat\d}} D^+_{\hat\a} D^+_{\hat\b}
  D^+_{\hat\g} D^+_{\hat\d} \cD^{--}L=0\;.
\end{equation}
In the flat case this is a direct consequence of \re{lincon}, but in the
curved case the algebra of the covariant derivatives is complicated by the
presence of torsion. Let us try to push all the spinor derivatives in
\re{compc} through $\cD^{--}$ until they reach $L$. This leads to the
following compatibility condition, which is the counterpart of the
four-dimensional constraint \re{consistency1}:
\begin{eqnarray}\label{incons}
 && \ve^{{\hat\a}{\hat\b}{\hat\g}{\hat\d}} [\cD^{--} D^+_{\hat\a}
   D^+_{\hat\b} D^+_{\hat\g} D^+_{\hat\d} L - 4 \cD^-_{\hat\a} D^+_{\hat\b}
   D^+_{\hat\g} D^+_{\hat\d} L + 12{\rm i}\, \cD_{{\hat\a}{\hat\b}}
   D^+_{\hat\g} D^+_{\hat\d} L]  \nonumber \\
 && 24{\rm i}\, T^{+{\hat\a}}{}_{{\hat\b}{\hat\g}} \cD^{{\hat\b}{\hat\g}}
   D^+_{\hat\a} L - 6{\rm i}\, ( D^+_{\hat\a} T^{+{\hat\a}}{}_{{\hat\b}
   {\hat\g}} - T^{+{\hat\a}}{}_{{\hat\a}{\hat\d}} T^{+{\hat\d}}{}_{{\hat\b}
   {\hat\g}})\cD^{{\hat\b}{\hat\g}} L = 0\;.
\end{eqnarray}
It is now obvious that keeping \re{lincon} in its flat form leads to an
inconsistency. This means that we have to modify \re{lincon} in such a way
that \re{incons} becomes a corollary. The necessary modification is
suggested by the comparison of the term
$\ve^{{\hat\a}{\hat\b}{\hat\g}{\hat\d}} \cD_{{\hat\a}{\hat\b}} D^+_{\hat\g}
D^+_{\hat\d} L$ with the torsion terms in \re{incons}:
\begin{equation}\label{newlincon}
  D^+_{\hat\a} D^+_{\hat\b} L = - T^{+{\hat\g}}{}_{{\hat\a}{\hat\b}}
  D^+_{\hat\g} L + {1\over 4}( D^+_{\hat\g} T^{+{\hat\g}}{}_{{\hat\a}
  {\hat\b}} - T^{+{\hat\g}}{}_{{\hat\g}{\hat\d}} T^{+{\hat\d}}{}_{{\hat\a}
  {\hat\b}}) L\;.
\end{equation}
Inserting this into the term $\ve^{{\hat\a}{\hat\b}{\hat\g}{\hat\d}}
\cD_{{\hat\a}{\hat\b}} D^+_{\hat\g} D^+_{\hat\d} L$ leads to the cancellation
of the torsion terms already present in \re{incons}, but produces two new
ones:
\begin{equation}\label{longeq}
  -24{\rm i}\, \cD^{{\hat\a}{\hat\b}} T^{+{\hat\g}}{}_{{\hat\a}{\hat\b}}
  D^+_{\hat\g} L + 6{\rm i}\, \cD^{{\hat\a}{\hat\b}} ( D^+_{\hat\g}
  T^{+{\hat\g}}{}_{{\hat\a}{\hat\b}} - T^{+{\hat\g}}{}_{{\hat\g}{\hat\d}}
  T^{+{\hat\d}}{}_{{\hat\a}{\hat\b}}) L\;.
\end{equation}
The only way to get rid of these new terms is to impose the following {\it
constraint on the torsion}:
\begin{equation}\label{torsconst}
  \cD^{{\hat\a}{\hat\b}}T^{+{\hat\g}}{}_{{\hat\a}{\hat\b}} = 0\;.
\end{equation}
Hitting \re{torsconst} with $D^+_{\hat\g}$, we easily see that the second
term in \re{longeq} vanishes as a corollary.

We have not yet finished checking the consistency conditions. The modified
constraint \re{newlincon} requires in its own turn a consistency test:
hitting its right-hand side with $D^+_{\hat\g}$ should produce a result
totally antisymmetric in ${\hat\a},{\hat\b},{\hat\g}$, like the left-hand
side. In fact, a simple calculation shows that the result is zero as a
corollary of the Bianchi identity \re{traceless}:
\begin{equation}
   D^+_{\hat\g} D^+_{\hat\a} D^+_{\hat\b} L = 0\;.
\end{equation}
As a direct consequence, the remaining three- and four spinor derivative
terms in \re{incons} vanish, so the consistency of the new constraint is
fully established. Note that in the process we have derived a condition on
the torsion, eq.\ \re{torsconst}. Written out in terms of the prepotentials
$H^{--\hm}$, this condition reads
\begin{equation}\label{torsconstprep}
  \cD^{{\hat\a}{\hat\b}} T^{+{\hat\g}}{}_{{\hat\a}{\hat\b}} =
  - {\rm i}\, (D^+)^{3{\hat\g}}\pa_\hm H^{--\hm}
  - {1\over 8} (D^+)^{3{\hat\g}} H^{--\hn} e_\hn^{{\hat\a}{\hat\b}}
  D^+_{\hat\a} D^+_{\hat\b} \pa_\hm H^{--\hm} = 0\;.
\end{equation}
One way to satisfy it is to put $\pa_\hm H^{--\hm}=0$ and we shall see below
that precisely this happens upon dimensional reduction to four dimensions.

\subsubsection{Four-dimensional formulation}

The framework developed in the preceding subsection allows us to obtain the
constraints of the linear VT multiplet in the presence of gauged central
charges by straightforward dimensional reduction of the constraints for the
self-dual tensor multiplet from six to four dimensions. Before doing this, we
must stress the important difference between the VT superfield $L$ and all
other superfields we are considering here (prepotentials $H$, sechsbein $e$,
torsion $T$). While the latter ceases to depend on the central charge
coordinates $x^{5,6}$ upon dimensional reduction (thus giving rise to
ordinary supermultiplets), $L$ keeps its dependence on $x^5$, but not on
$x^6$:
\begin{equation}\label{5not6}
  L = L(x^m,x^{5},\q,u)\;.
\end{equation}
To see why we should only allow for one non-trivial central charge, we can
look, for instance, at the six-dimensional equation of motion for the scalar
$\o(x)$ in (\ref{eqmot}). In four-dimensional notation it becomes
\begin{equation}
  (\pa^2_5 + \pa^2_6)\o = \Box\o\;.
\end{equation}
This is not an equation of motion any more, but rather an equation allowing
to relate the dependence on the extra central-charge coordinates to that
on $x^m$. It is then clear that by putting, e.g., $\pa_6\o = 0$ we can
completely fix the dependence on the remaining central-charge coordinate
$x^5$, whereas keeping both $x^5$ and $x^6$ would leave a functional
freedom in some combination of those coordinates. Thus, the VT multiplet
carries only one central charge and is inert under the action of the
second one.

All we need to do now is to compute the torsion $T^+$. The key ingredient in
this is the sechsbein matrix $e^\hm_{\hat\a\hat\b}$:
\begin{equation}
  e^\hm_{\hat\a\hat\b}=
 \left(
  \begin{array}{ccc}
    e^m_{\a\bd}=\s^m_{\a\bd} & e^5_{\a\bd} & e^6_{\a\bd} \\
    e^m_{\a\b}=0 & e^5_{\a\b} = {\rm i}\ve_{\a\b}\bar\cZ & e^6_{\a\b}
    = {\rm i}\ve_{\a\b}\bar\cY\\
    e^m_{\ad\bd}=0 & e^5_{\ad\bd} = -{\rm i}\ve_{\ad\bd}\cZ & e^6_{\ad\bd}
    = - {\rm i}\ve_{\ad\bd}\cY\\
  \end{array}
 \right)
\end{equation}
Note that the elements $e^{5,6}_{\a\bd}=-{{\rm i}\over 2}D^+_\a \bar D^+_\bd
H^{--5,6}$ are not gauge invariant, but they will not show up in any of the
expressions below. The elements of the inverse matrix $e_\hm^{\hat\a\hat\b}$
relevant to our calculation are
\begin{eqnarray}
  && e_{5\a\bd}=0\;, \quad e_{5\a\b} = {2{\rm i}\bar\cY \over \bar\cZ\cY -
     \cZ\bar\cY}\ve_{\a\b}\;, \quad e_{5\ad\bd}= {2{\rm i}\cY\over \bar\cZ\cY
     - \cZ\bar\cY}\ve_{\ad\bd}\;; \nonumber\\
  && e_{6\a\bd}=0\;, \quad e_{6\a\b} = {-2{\rm i}\bar\cZ\over \bar\cZ\cY -
     \cZ\bar\cY}\ve_{\a\b}\;, \quad e_{6\ad\bd} = {-2{\rm i}\cZ\over \bar\cZ
     \cY -\cZ\bar\cY}\ve_{\ad\bd}\;.
\end{eqnarray}
Then the non-vanishing torsion components are
\begin{equation}
  T^{+\a}{}_{\b\g} = {\bar\cZ D^{+\a}\cY - \bar\cY D^{+\a}\cZ \over
   \bar\cZ\cY -\cZ\bar\cY}\ve_{\b\g}\;, \qquad
  T^{+\ad}{}_{\b\g} = {\bar\cY \bar D^{+\ad}\bar\cZ  - \bar\cZ \bar
   D^{+\ad}\bar\cY \over \bar\cZ\cY -\cZ\bar\cY}\ve_{\b\g}
\end{equation}
and complex conjugates.

Now we have to put all of this into the defining constraint \re{newlincon}.
We should not forget that this constraint was only consistent under the
condition \re{torsconst} on the torsion. From \re{torsconstprep} we see
that the condition is satisfied if $\pa_\hm H^{--\hm}=0$, but this is
precisely what happens in our dimensional reduction scheme:
\begin{equation}
  \pa_\hm H^{--\hm} = \pa_m (-2{\rm i}\q^-\s^m\bar\q^-) +\pa_5 H^{--5} +
  \pa_6 H^{--6}=0\;.
\end{equation}
Thus, we find the final form of the constraints of the linear VT multiplet
in the presence of gauged central charges:
\begin{eqnarray}
  D^+_\a\bar D^+_\bd L &=& 0\;, \label{gaugcon1}  \\
  D^+ D^+ L  &=& {2\over \bar\cZ\cY -\cZ\bar\cY}
  \Big[ (\bar\cY D^{+\a}\cZ - \bar\cZ D^{+\a}\cY )D^+_\a L +
  (\bar\cY \bar D^+_{\dot\a} \bar\cZ  - \bar\cZ \bar D^+_{\dot\a} \bar\cY )
  \bar D^{+\dot\a}L \nonumber \\
  && + {1\over 2}(\bar\cY D^+D^+\cZ - \bar\cZ D^+D^+\cY)L \Big]\;.
     \label{gaugcon2}
\end{eqnarray}
They are the $\l$-frame form of those given in subsection \ref{scinv},
eq.\ \re{lsc2}.

\subsection{Coupling to a super-Yang-Mills multiplet} 

\subsubsection{SYM in six dimensions}

In six dimensions the SYM multiplet is described by an analytic prepotential
$V^{++}$ (the analog of $H^{++5,6}$ in the case of the gauged (abelian)
central charge) which serves as a gauge connection for the harmonic
derivative:
\begin{equation}
  \cD^{++} = D^{++} + {\rm i}V^{++}\;.
\end{equation}
It undergoes the gauge transformations
\begin{equation}
  \d V^{++} = -\cD^{++}\Lambda = -D^{++}\Lambda - {\rm i}[V^{++},\Lambda]
\end{equation}
with an analytic gauge parameter $\Lambda(x,\q^+,u)$. Next, going through
steps similar to those in the case of the prepotentials $H^{++}$ and
$H^{--}$ above, we first define the connection $V^{--}$ related to
$V^{++}$ by the constraint
\begin{equation}\label{harmcurv}
  D^{++}V^{--} - D^{--}V^{++} + {\rm i}[V^{++},V^{--}] = 0
\end{equation}
and from it we construct the field strength:
\begin{equation}\label{fstr}
  \cW^{+{\hat\a}} = {1\over 2}(D^+)^{3{\hat\a}}V^{--}\;.
\end{equation}
Let us verify that this expression is gauge invariant if the vielbeins $H$
are flat (\ref{flatH}):
\begin{equation}\label{isginv}
  \d\cW^{+{\hat\a}} = {1\over 12} \ve^{{\hat\a}{\hat\b}{\hat\g}{\hat\d}}
  D^+_{\hat\b} D^+_{\hat\g} D^+_{\hat\d} \cD^{--}\Lambda =
  - {1\over 12} \ve^{{\hat\a}{\hat\b}{\hat\g}{\hat\d}} D^+_{\hat\b}
  \cD_{{\hat\g}{\hat\d}} \Lambda = 0
\end{equation}
since $[D^+_{\hat\b},\cD_{{\hat\g}{\hat\d}}]=0$ and $D^+_{\hat\b}\Lambda = 0$.

The field strength \re{fstr} satisfies the harmonic constraint
\begin{equation}\label{hconfstr}
  \cD^{++}\cW^{+{\hat\a}} = 0
\end{equation}
(the proof is similar to \re{isginv} and makes use of \re{harmcurv}), as
well as the obvious spinor one
\begin{equation}\label{sconfstr}
  D^+_{({\hat\b}}\cW^{+{\hat\a})} = 0\;.
\end{equation}

This very simple picture is distorted when we take into account the torsion
introduced in subsection \ref{subsec}. First of all, the expression
\re{fstr} is not gauge invariant any more:
\begin{equation}
  \d (\ve^{{\hat\a}{\hat\b}{\hat\g}{\hat\d}} D^+_{\hat\b} D^+_{\hat\g}
  D^+_{\hat\d} V^{--}) = \ve^{{\hat\a}{\hat\b}{\hat\g}{\hat\d}}
  D^+_{\hat\b} \cD_{{\hat\g}{\hat\d}} \Lambda = -3 T^{+{\hat\a}}{}_{{\hat\b}
  {\hat\g}} \cD^{{\hat\b}{\hat\g}} \Lambda\;.
\end{equation}
This can be compensated for by adding the term
\begin{equation}
  \d (T^{+{\hat\a}{\hat\b}{\hat\g}} D^+_{\hat\b} D^+_{\hat\g} V^{--})
  = - T^{+{\hat\a}}{}_{{\hat\b}{\hat\g}} \cD^{{\hat\b}{\hat\g}} \Lambda\;.
\end{equation}
Thus, the combination
\begin{equation}\label{modfstr}
  \cW^{+{\hat\a}} = {1\over 2}(D^+)^{3{\hat\a}}V^{--} -
  {1\over 4} T^{+{\hat\a}{\hat\b}{\hat\g}} D^+_{\hat\b} D^+_{\hat\g} V^{--}
\end{equation}
is gauge invariant and becomes the torsion-modified expression of the SYM
field strength. This new field strength \re{modfstr} satisfies the old
harmonic constraint \re{hconfstr}, but the spinor one \re{sconfstr} is
modified by the torsion:
\begin{equation}\label{sconfstrmod}
  D^+_{({\hat\b}}\cW^{+{\hat\a})} -
T^{+({\hat\a}}{}_{{\hat\b}){\hat\g}} \cW^{+{\hat\g}}= 0\;.
\end{equation}

Although the construction outlined so far is sufficient to describe SYM in
six dimensions, we point out that we may further modify the expression
\re{modfstr} (this will turn out useful when coupling the VT multiplet to
SYM). The determinant of the sechsbein matrix
$e=\det(e^\hm_{{\hat\a}{\hat\b}})$ is a density, i.e.\ a quantity
transforming homogeneously under the diffeomorphism group:
\begin{equation}
  \d (\ln e) = -{1\over 4} \d (e^\hm_{{\hat\a}{\hat\b}})
  e_\hm^{{\hat\a}{\hat\b}} = {{\rm i}\over 8}D^+_{\hat\a} D^+_{\hat\b}
  \cD^{--}\l^\hm e_\hm^{{\hat\a}{\hat\b}} = -{1\over 4}
  e^\hn_{{\hat\a}{\hat\b}}\pa_\hn \l^\hm e_\hm^{{\hat\a}{\hat\b}}
  = \pa_\hm\l^\hm
\end{equation}
(note that $\pa_\hm\l^\hm =0$ after dimensional reduction, so $e$ is
invariant in four dimensions). Then it is clear that its spinor derivative
$D^+_{\hat\a} \ln e$ is invariant and as such must be related to the torsion
tensor. Indeed,
\begin{equation}
  D^+_{\hat\a} \ln e = -{1\over 4}D^+_{\hat\a} e^\hm_{{\hat\b}{\hat\g}}
  e_\hm^{{\hat\b}{\hat\g}} = {{\rm i}\over 8} D^+_{\hat\a} D^+_{\hat\b}
  D^+_{\hat\g} H^{--\hm} e_\hm^{{\hat\b}{\hat\g}} =
  T^{+{\hat\b}}{}_{{\hat\a}{\hat\b}}\;.
\end{equation}
All this allows us to redefine $\cW^{+{\hat\a}}$ by multiplying it by the
density $e$, for instance,
\begin{equation}\label{hatW}
  \hat \cW^{+{\hat\a}} = e^{-1/2}\cW^{+{\hat\a}}\;.
\end{equation}
Then we find that the constraint \re{sconfstrmod} is further modified:
\begin{equation}\label{sconfstrmodd}
  D^+_{({\hat\b}} \hat\cW^{+{\hat\a})} - T^{+({\hat\a}}{}_{{\hat\b}){\hat\g}}
  \hat\cW^{+{\hat\g}} - {1\over 2} T^{+{\hat\g}}{}_{{\hat\g}({\hat\a}}
  \hat\cW^{+{\hat\b})}= 0\;.
\end{equation}
Note also that the harmonic constraint \re{hconfstr} acquires a new term:
\begin{equation}
  (\cD^{++} +{1\over 2}\pa_\hm H^{++\hm}) \hat\cW^{+{\hat\b}}= 0\;,
\end{equation}
which serves as a connection for the density transformations (once again,
this term vanishes upon dimensional reduction).

\subsubsection{Chern-Simons coupling}

The so-called Chern-Simons coupling of the VT multiplet to a SYM multiplet is
realized in an obvious way in six-dimensional flat superspace. There its
analog is the coupling of the self-dual tensor multiplet to SYM \cite{BSS}:
\begin{equation}\label{YMcoupfl}
  D^+_{\hat\a} D^+_{\hat\b} L = -{1\over 2} \ve_{{\hat\a}{\hat\b}{\hat\g}
  {\hat\d}}\, {\rm tr}\, (\cW^{+{\hat\g}} \cW^{+{\hat\d}})\;.
\end{equation}
Note that the superfield $L$ is a singlet with respect to the Yang-Mills
group. The consistency condition for (\ref{YMcoupfl}) is obtained by
hitting it with $D^+_{\hat\g}$ and using the flat superspace SYM
constraint \re{sconfstr}:
\begin{equation}\label{rhseq}
  D^+_{\hat\g} D^+_{\hat\a} D^+_{\hat\b} L = -
  {1\over 4} \ve_{{\hat\a}{\hat\b}{\hat\g}{\hat\d}}\, {\rm tr}\,
  ((D^+_{\hat\r} \cW^{+\hat\r}) \cW^{+{\hat\d}})\;.
\end{equation}
Clearly, the right-hand side of eq.\ \ref{rhseq} is totally antisymmetric in
${\hat\a},{\hat\b},{\hat\g}$, just like the left-hand side.

The problem now is to properly covariantize the coupling \re{YMcoupfl} in
the presence of torsion. This is not too difficult. We just have to put
together \re{newlincon} with \re{YMcoupfl} and replace $\cW^{+{\hat\a}}$
by $\hat\cW^{+{\hat\a}}$ defined in \re{hatW}:
\begin{equation}\label{puttog}
  D^+_{\hat\a} D^+_{\hat\b} L = - T^{+{\hat\g}}{}_{{\hat\a}{\hat\b}}
  D^+_{\hat\g} L + {1\over 4} (D^+_{\hat\g} T^{+{\hat\g}}{}_{{\hat\a}
  {\hat\b}} - T^{+{\hat\g}}{}_{{\hat\g}{\hat\d}} T^{+{\hat\d}}{}_{{\hat\a}
  {\hat\b}}) L - {1\over 2} \ve_{{\hat\a}{\hat\b}{\hat\g}{\hat\d}}\,
  {\rm tr}\, (\hat\cW^{+{\hat\g}} \hat\cW^{+{\hat\d}})\;.
\end{equation}
Repeating the consistency check \re{rhseq}, we find the following new terms:
\begin{equation}
  T^{+\hat\r}{}_{{\hat\a}{\hat\b}} \ve_{{\hat\g}\hat\r\hat\k\hat\s}\,
  {\rm tr}\, (\hat\cW^{+\hat\k} \hat\cW^{+\hat\s}) + 2 \ve_{{\hat\a}{\hat\b}
  \hat\k\hat\s}\, {\rm tr}\, ((D^+_{\hat\g} \hat\cW^{+\hat\k}
  \hat\cW^{+\hat\s})
\end{equation}
which should be totally antisymmetric in ${\hat\a},{\hat\b},{\hat\g}$. It is
a matter of a straightforward calculation to show that this is true as a
consequence of the constraint \re{sconfstrmodd} satisfied by the
density-modified field strength $\hat\cW^{+{\hat\a}}$. In addition, the
constraint \re{puttog} passes the consistency check \re{incons}.

\subsubsection{Four-dimensional formulation}

Having achieved a consistent coupling in six dimensions, we just need
to carry out the dimensional reduction. Here is a sketch. First, the
determinant of the sechsbein is given in terms of the field strengths
for the gauged central charges:
\begin{equation}
  e = -4{\rm i}\, (\bar\cZ\cY -\cZ\bar\cY)\;.
\end{equation}
Next, the expression \re{modfstr} for the SYM field strength is shown to
have two equivalent forms:
\begin{equation}
  \cW^{+\a} = e D^{+\a}\left({\bar\cZ\cW -\cZ\bar\cW \over e\bar\cZ}\right)
   = e D^{+\a}\left({\bar\cY\cW -\cY\bar\cW \over e\bar\cY}\right)\;.
\end{equation}
According to \re{puttog}, this leads to the following modification of the
constraint \re{gaugcon1}:
\begin{equation}
  D^+_\a \bar D^+_{\bd} L = D^+_\a
  \bar D^+_{\bd}\; {\rm tr}\,
\left({(\bar\cZ\cW -\cZ\bar\cW)^2 \over e\cZ\bar\cZ}\right)\;,
\end{equation}
which immediately suggests to redefine $L$:
\begin{equation}
  {\Bbb L} = L - {\rm tr}\,{(\bar\cZ\cW -\cZ\bar\cW)^2 \over e\cZ\bar\cZ}
\end{equation}
in order to recover the simple form \re{gaugcon1} of the constraint.
Finally, the other constraint \re{gaugcon2} turns into the second
constraint in \re{lsc4}.

\sect{Conclusion}

In the present paper we have developed the harmonic superspace setting for
general $N=2$ rigid supersymmetric theories with gauged central charge. We have
constructed the superfield formulations for both the linear and nonlinear VT
multiplets with gauged central charge and described their Chern-Simons
couplings to $N=2$ vector multiplets. The six-dimensional origin of the linear
VT multiplet and its Chern-Simons couplings has been explained. Note that the
constraints describing the nonlinear VT multiplet can also be written in a
six-dimensional notation but using a constant tensor which breaks the
six-dimensional Lorentz symmetry down to its four-dimensional part.

The VT multiplet superfield formulations developed in this paper allow one to
couple the VT multiplet to $N=2$ superfield supergravity according to the
general rules given in \cite{gns,gios2}. To do this one should find consistent
curved superspace extensions of the VT multiplet constraints presented above.

The analysis of this paper was restricted to the case of a single VT multiplet
but the whole consideration can be readily applied to theories with several VT
multiplets. There remains, however, an interesting problem whether there exist
consistent theories with several VT multiplets, linear or/and nonlinear ones,
coupled to each other. It would also be interesting to study the gauged central
charge version of the `new' nonlinear VT multiplet proposed in \cite{is}.

\vspace{1cm}

\noindent
{\bf Acknowledgments.}
EI is grateful to P. Sorba for hospitality at LAPTH during the final stage
of this work. SK is grateful to J. Louis for kind hospitality extended to
him at Martin-Luther-Universit\"at Halle-Wittenberg. EI and SK acknowledge
a partial support from INTAS grant, INTAS-96-0308 and from RFBR-DFG grant,
project No 96-0200180. The work of EI was supported in part by RFBR grant,
project No 96-02-17634, by INTAS grants, INTAS-93-127-ext and INTAS-96-0538
and by PICS grant No 593. The work of SK was partially supported by the
Alexander von Humboldt Foundation and RFBR grant, project No 96-02-1607.


\begin{thebibliography}{99}
\bibitem{z} C. Zachos, Phys. Lett. {\bf 76B} (1978) 329.
\bibitem{dvv} B. de Wit, J.W. van Holten and A. Van Proeyen,
Nucl. Phys. {\bf B 167} (1980) 186 (E: ibid. {\bf B 172} (1980) 543);
Phys. Lett. {\bf 95B} (1980) 51;
Nucl. Phys. {\bf B 184} (1981) 77 (E: ibid. {\bf B 222} (1983) 516).
\bibitem{wkll} B. de Wit, V. Kaplunovsky, J. Louis and D. L\"ust,
Nucl. Phys. {\bf B 451} (1995) 53.
\bibitem{ssw} M.F. Sohnius, K.S. Stelle and P. West, Phys. Lett.
{\bf 92B} (1980) 123; Nucl. Phys. {\bf B 173} (1980) 127.
\bibitem{claus1} P. Claus, B. de Wit, M. Faux, B. Kleijn, R. Siebelink
and P. Termonia, Phys. Lett. {\bf 373B} (1996) 81.
\bibitem{claus2} P. Claus, B. de Wit, M. Faux and P. Termonia,
Nucl. Phys. {\bf B 491} (1997) 201.
\bibitem{claus3} P. Claus, B. de Wit, M. Faux, B. Kleijn, R. Siebelink and
P. Termonia, Nucl. Phys. {\bf B 512} (1998) 148.
\bibitem{how} A. Hindawi, B.A. Ovrut and D. Waldram,
Phys. Lett. {\bf 392B} (1997) 85.
\bibitem{ghh} R. Grimm, M. Hasler and C. Herrmann, {\it The $N=2$
vector-tensor multiplet, central charge superspace and Chern-Simons
couplings}, hep-th/9706108.
\bibitem{bho} I. Buchbinder, A. Hindawi and B.A. Ovrut,
Phys. Lett. {\bf 413B} (1997) 79.
\bibitem{dkt} N. Dragon, S.M. Kuzenko and U. Theis, {\it The
vector-tensor multiplet in harmonic superspace}, hep-th/9706169
(to appear in Eur. Phys. J. {\bf C}).
\bibitem{gikos} A. Galperin, E. Ivanov, S. Kalitzin, V. Ogievetsky and
E. Sokatchev, Class. Quant. Grav. {\bf 1} (1984) 469.
\bibitem{dk} N. Dragon and S. M. Kuzenko, Phys. Lett. {\bf 420B} (1998) 64.
\bibitem{is} E. Ivanov and E. Sokatchev, {\it On non-linear
superfield versions of the vector-tensor multiplet}, hep-th/9711038
(to appear in Phys. Lett. {\bf B}).
\bibitem{dt} N. Dragon and U. Theis, {\it Gauging the central charge},
hep-th/9711025.
\bibitem{fs} P. Fayet, Nucl. Phys. {\bf B 113} (1976) 135.
\bibitem{fs1} M.F. Sohnius, Nucl. Phys. {\bf B 138} (1978) 109.
\bibitem{gsw} R. Grimm, M. Sohnius and J. Wess, Nucl. Phys.
{\bf B 133} (1978) 275.
\bibitem{zup} B. Zupnik, Phys. Lett. {\bf B 183} (1987) 175.
\bibitem{gio} A. Galperin, E. Ivanov and V. Ogievetsky,
Nucl. Phys. {\bf B 282} (1987) 74.
\bibitem{bk} I.L. Buchbinder and S.M. Kuzenko, Class. Quant. Grav.
{\bf 14} (1997) L157.
\bibitem{ikz} E.A. Ivanov, S.V. Ketov and B.M. Zupnik, Nucl. Phys.
{\bf B 509} (1997) 53.
\bibitem{SSw} J. Scherk and J. Schwarz, Nucl. Phys. {\bf B 153} (1979) 61.
\bibitem{bd} F. Brandt and N. Dragon, {\it Nonpolynomial gauge invariant
interactions of 1-form and 2-form gauge potentials}, hep-th/9709021.
\bibitem{HST}  P.S. Howe, G. Sierra and P.K. Townsend,
Nucl. Phys. {\bf B 221} (1983) 331.
\bibitem{HSW} P.S. Howe, K.S. Stelle and P.C. West,
Class. Quant. Grav. {\bf 2} (1985) 815.
\bibitem{zupnik} B.M. Zupnik, Sov. J. Nucl. Phys. {\bf 44} (1986) 512.
\bibitem{BPSez} E. Bergshoeff, E. Sezgin and A. Van Proeyen,
Nucl. Phys. {\bf B 264} (1986) 653.
\bibitem{Sok} E. Sokatchev, Class. Quant. Grav. {\bf 5} (1988) 1459.
\bibitem{BSS} E. Bergshoeff, E. Sezgin and E. Sokatchev,
Class. Quant. Grav. {\bf 13} (1996) 2875.
\bibitem{gns} A. Galperin, Nguyen Ahn Ky and E. Sokatchev,
Class. Quant. Grav. {\bf 4} (1987) 1235.
\bibitem{gios2} A. Galperin, E. Ivanov, V. Ogievetsky and
E. Sokatchev, Class. Quant. Grav. {\bf 4} (1987) 1255.
\end{thebibliography}
\end{document}